\documentclass[default,iicol]{sn-jnl}

\usepackage{graphicx}%
\usepackage{multirow}%
\usepackage{amsmath,amssymb,amsfonts}%
\usepackage{amsthm}%
\usepackage{mathrsfs}%
\usepackage[title]{appendix}%
\usepackage{xcolor}%
\usepackage{textcomp}%
\usepackage{manyfoot}%
\usepackage{booktabs}%
\usepackage{algorithm}%
\usepackage{algorithmicx}%
\usepackage{algpseudocode}%
\usepackage{listings}%
\usepackage{siunitx}
\usepackage[normalem]{ulem}
\usepackage{anyfontsize}
\usepackage{float}
\usepackage{bookmark}
\usepackage{tabularx}

\begin{document}
\vspace{-2cm}
\title[Article Title]{A unified realization of electrical quantities from the quantum International System of Units} 

\author[1,2]{\fnm{Linsey K.} \sur{Rodenbach}}
\equalcont{These authors contributed equally to this work.}
\author*[3]{\fnm{Jason M.} \sur{Underwood}}
\equalcont{These authors contributed equally to this work.}\email{jason.underwood@nist.gov}
\author[3,4]{\fnm{Ngoc Thanh Mai} \sur{Tran}}
\author[3]{\fnm{Alireza R.} \sur{Panna}}
\author[2,5]{\fnm{Molly P.} \sur{Andersen}}
\author[3,4]{\fnm{Zachary S.} \sur{Barcikowski}}
\author[3]{\fnm{Shamith U.} \sur{Payagala}}
\author[6]{\fnm{Peng} \sur{Zhang}}
\author[6]{\fnm{Lixuan} \sur{Tai}}
\author[6]{\fnm{Kang L.} \sur{Wang}}
\author[3]{\fnm{Dean G.} \sur{Jarrett}}
\author[3]{\fnm{Randolph E.} \sur{Elmquist}}
\author[3]{\fnm{David B.} \sur{Newell}}
\author[3]{\fnm{Albert F.} \sur{Rigosi}}
\author*[1,2]{\fnm{David} \sur{Goldhaber-Gordon}}\email{goldhaber-gordon@stanford.edu }

\affil[1]{\orgdiv{Department of Physics}, \orgname{Stanford University}, \orgaddress{\city{Stanford}, \state{California}, \postcode{94305}, \country{USA}}}

\affil[2]{\orgdiv{Stanford Institute for Materials and Energy Sciences}, \orgname{SLAC National Accelerator Laboratory}, \orgaddress{\street{2575 Sand Hill Road}, \city{Menlo Park}, \state{California}, \postcode{94025},\country{USA}}}

\affil[3]{\orgdiv{Physical Measurement Laboratory}, \orgname{National Institute of Standards and Technology (NIST)}, \orgaddress{\city{Gaithersburg}, \state{Maryland}, \postcode{20899}, \country{USA}}}

\affil[4]{\orgdiv{Joint Quantum Institute}, \orgname{University of Maryland}, \orgaddress{\city{College Park}, \state{Maryland}, \postcode{20742}, \country{USA}}}

\affil[5]{\orgdiv{Department of Materials Science and Engineering}, \orgname{Stanford University}, \orgaddress{\city{Stanford}, \state{California}, \postcode{94305}, \country{USA}}}

\affil[6]{\orgdiv{Department of Electrical and Computer Engineering}, \orgname{University of California, Los Angeles}, \orgaddress{\city{Los Angeles}, \state{California}, \postcode{90095}, \country{USA}}}

\abstract{
\unboldmath{In the revised International System of Units (SI), the ohm and the volt are realized from the von Klitzing constant and the Josephson constant, and a practical realization of the ampere is possible by applying Ohm’s law directly to the quantum Hall and Josephson effects. As a result, it is possible to create an instrument capable of realizing all three primary electrical units, but the development of such a system remains challenging. Here we report a unified realization of the volt, ohm, and ampere by integrating a quantum anomalous Hall resistor (QAHR) and a programmable Josephson voltage standard (PJVS) in a single cryostat. Our system has a quantum voltage output that ranges from 0.24~mV to 6.5~mV with combined relative uncertainties down to 3~$\mu$V/V. The QAHR provides a realization of the ohm at zero magnetic field with uncertainties near 1~$\mu\Omega$/$\Omega$. We use the QAHR to convert a longitudinal current to a quantized Hall voltage and then directly compare that against the PJVS to realize the ampere. We determine currents in the range of 9.33~nA to 252~nA, and our lowest uncertainty is 4.3~$\mu$A/A at 83.9~nA. For other current values, a systematic error that ranges from -10~$\mu$A/A to -30~$\mu$A/A is present due to the imperfect isolation of the PJVS microwave bias.}
}
\keywords{Quantum metrology, quantum Hall, Topological insulators, Superconducting devices, Techniques and instrumentation, Standards}


\maketitle
\section*{1 Main}\label{section: main}

In 2019, the International System of Units (SI) was redefined to link SI units to physical constants in nature~\cite{mills2005,milton2014,davis2019}. Under the new SI, units are defined solely by their relation to fixed-value fundamental constants. This redefinition removed from unit definitions all volatile artifacts (objects such as a 100~$\Omega$ resistor or 1~kg mass) and instead endorsed practical \emph{realizations} of SI units via quantum phenomena~\cite{davis2019,poirier2019}.

Before the 2019 redefinition, the ohm ($\Omega$) and the volt (V) were referenced by precisely measuring the quantum Hall and Josephson effects, respectively. However, these quantum effects were not realizations of their respective SI-derived units because of the stringent definitions for the SI base units the ampere (A) and the kilogram (kg) (recall the unit relations: V~=~kg~m$^2$~s$^{-3}$~A$^{-1}$ and $\Omega =$~kg~m$^2$~s$^{-3}$~A$^{-2}$). Even so, precise measurements of those quantum effects were widely used to maintain artifact standards~\cite{taylor1989} for resistance~\cite{jeckelmann2001,tzalenchuk2010,schopfer2013,ribeiro2015, rigosi2019} and voltage~\cite{Bloch1970,Fulton1973,Clothier1989} . 
These artifact standards were in turn used to maintain electrical current~\cite{poirier2019} artifact standards, via the unit relation A = V/$\Omega$ (Ohm's law). This convoluted approach made traceability of measurements burdensome.

Under the modern SI, the ohm and the volt are realized from the von Klitzing constant ($R_K=h/e^2$) and the Josephson constant ($K_J=2e/h$), and the exactly defined values of the elementary charge ($e$) and the Planck constant ($h$). Likewise, a practical realization of the ampere is now possible by applying Ohm’s law directly to the quantum Hall and Josephson effects~\cite{Mohr2007,Keller2008,Scherer_2012,davis2019}. Such harmonization of the primary electrical units suggests that it should be possible to develop an instrument capable of realizing all three. Indeed, one early example involved the on-chip integration of a quantum Hall resistance standard (QHRS) and a current pump based on single electron transport (SET) to produce a quantum realization of the volt, without any superconductors~\cite{sqvs2012}. A theoretical analysis of the device suggests a potential minimum uncertainty of about $10^{-8}$~$\mu$A/A for the current pump. However, the output voltage of the device was 10~$\mu$V and the combined measurement uncertainties were around 1\%.

More recently, a programmable quantum current generator (PQCG)~\cite{Brun-Picard_2016,Djordjevic_2025} has been developed that combines a QHRS with a programmable Josephson voltage standard (PJVS) and a cryogenic current comparator (CCC)~\cite{sullivan1974low,williams2011}. The approach provides a realization of the ampere from 1~$\mu$A to 10~mA with relative uncertainties of $\leq 10^{-8}$~\cite{Brun-Picard_2016,Djordjevic_2025}. However, due to differing magnetic field requirements for the QHRS and PJVS, the two quantum standards were operated in separate cryostats. Under the revised SI, the ampere can also be realized by using a PJVS in combination with a quantum Hall resistance array standard (QHRAS), which uses serial-parallel combinations of quantum Hall resistors to achieve quantized resistances well beyond $R_K$. Laboratory-scale PJVS–QHRAS systems have demonstrated quantum-based generation and sensing of current from 10~nA to 1~$\mu$A, with uncertainties from 4~$\mu$A/A to just under 0.1~$\mu$A/A, respectively~\cite{Chae_2020,Chae_2022}.

In this Article we report the construction of a unified realization of the SI volt, ohm, and ampere, based on the direct integration of a quantum anomalous Hall resistor (QAHR) and a PJVS within a single cryostat. The approach is possible due to the zero-field quantization of the Hall resistivity ($\rho_{yx} =R_{yx}= R_K$) of the quantum anomalous Hall (QAH) state~\cite{chen2010,yu2010,checkelsky2012}, which has been shown to be a viable alternative to traditional QHRSs~\cite{Fox2018,Gotz2018,rodenbach2022,okazaki2022,Patel2024}. We evaluate the accuracy of our integrated realization for the ohm and the volt through transfer standards to independent quantum standards: a gallium arsenide (GaAs)-based QHRS for the ohm and an ex-situ PJVS for the volt. The transfer standards for the ohm and volt have relative uncertainties of a few parts in $10^{6}$ or less.

For the ampere, we assess the accuracy through comparison to a separate, practical realization based on transfer standards and the QAHR. Our prototype has a quantum voltage output that ranges from 0.24~mV to 6.5~mV with combined relative uncertainties down to 3~$\mu$V/V. The co-located QAHR realizes the ohm at uncertainties near 1~$\mu\Omega$/$\Omega$. We use the QAHR to convert a longitudinal current to a quantized Hall voltage and then directly compare that against the PJVS to realize the ampere.

\section*{2 The unified quantum system}\label{section: design}
In our prototype, the PJVS and QAHR were co-located inside a dry $^3$He/$^4$He dilution refrigerator, with a base temperature of approximately 10~mK. The PJVS was mounted on the refrigerator's 4~K stage and covered by a copper radiation shield. The QAHR was mounted to a sample stage at the end of a cold finger below the mixing chamber plate. Direct current (DC) electrical connections between the QAHR, PJVS, and room temperature electronics were made through copper twisted pair wiring. Inside the cryostat, the twisted pairs were embedded within cryogenic woven loom cable, which was thermalized at the 60~K, 4~K, and mixing chamber temperature stages. 

\begin{figure}[htb]
   \centering \includegraphics[scale=1,keepaspectratio]{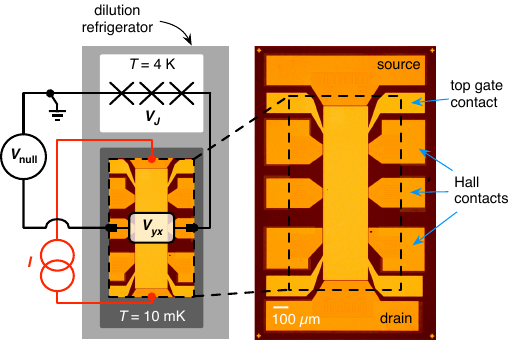}
       \caption{\textbf{Illustration of a current measurement with the unified system.} Simplified schematic for the direct realization of the ampere (left). A programmable Josephson voltage standard (PJVS, white background, denoted as $V_J$ in diagram), is mounted to the 4~K stage of a dilution refrigerator (light gray background, with $T$ denoting temperature). A quantum anomalous Hall resistor (QAHR, dark gray background, denoted by $V_{yx}$) is mounted to the mixing chamber of the dilution refrigerator. The measured base temperature of the mixing chamber stage was $T=10$~mK. The current source, denoted as $I$, biases the QAHR, producing the transverse voltage $V_{yx} = I R_{yx}$, where $R_{yx}$ is the quantized Hall resistance. The voltage difference $V_J-V_{yx}$ is measured by a nanovoltmeter (denoted by $V_\mathrm{null}$). If the system is biased such that $V_\mathrm{null}=0$, the current $I$ may be estimated from the number of Josephson junctions $N$ biased at frequency $f$ and the elementary charge $e$, according to $I=(e/2)Nf$. The nanovoltmeter and current source were external to the cryostat and at room temperature (approximately $T=296~K$). Also shown is an optical micrograph of the Cr-doped (Bi,Sb)$_2$Te$_3$ QAHR, showing device connections and scale (right).}
       \label{fig.1}
\end{figure}

A schematic depicting the conceptual design of the unified quantum instrument is shown in Fig.~\ref{fig.1} (left). The PJVS output voltage $V_J$ is placed in a series circuit with the transverse (Hall) voltage $V_{yx}$ of the QAHR and a nanovoltmeter (external to the cryostat). An additional pair of leads (not shown) runs directly from the PJVS to an external breakout box. These leads were used for performance verification of the PJVS and would, in a more mature realization, be used for voltage calibration of external equipment. 

The port used for connection of the nanovoltmeter was also used for evaluation of QAHR quantization with a separate CCC. This is possible because when the PJVS is inactive, the intervening wire resistance is small enough (a few ohms) and leakage resistance high enough (tens of G$\Omega$), that errors due to voltage division are negligible. As with the case for voltage, in a future, more robust design, this port could be used for calibration of external resistors against the QAHR through a resistance bridge. 

For realization of the ampere, our unified instrument operates by converting the externally applied current-under-test $I$ to a transverse voltage $V_{yx}$ via the dissipationless, topological state of the quantum anomalous Hall effect. The conversion factor is the quantized resistance $R_{yx} = R_K$. The Hall voltage $V_{yx}$ is then directly compared to a quantized, calculable voltage from an array of $N$ Josephson junctions phase locked to a frequency $f$: $V_J = N f K_J^{-1}$~\cite{Clothier1989, Stewart1968, Kautz1992, Kautz1995}. $V_J$ is tied to the high-potential side of the QAHR so that the potential difference $V_\mathrm{null} = V_J - V_{yx}$ can be measured by a nanovoltmeter at room temperature. The primary benefit of this arrangement is that it substantially reduces the accuracy requirements of the nanovoltmeter (see Supplementary Information section S4.1). 

If $f$ is tuned such that $V_\mathrm{null} = 0$ the current is then known to be
\begin{equation}
    \label{eq.Isource}
     I_\mathrm{direct} = \frac{N f}{K_J R_K} = \frac{e}{2}Nf.
\end{equation}
Here the subscript `direct' is used to indicate that this measurement is a `direct realization' of the ampere---that is, the measurement of $I$ is done through the integration of two quantum standards in a manner consistent with the revised SI. Generalizing to the case of nonzero null voltage and imperfect quantization of the QAHR, the current can still be determined according to,  
\begin{equation}
    \label{eq.Idirect}
    I_\mathrm{direct} = \frac{N f K_J^{-1} - V_\mathrm{null}}{R_{yx}(I,f)}, 
\end{equation}
where $R_{yx}(I,f) \simeq R_K$ is the \emph{measured} Hall resistance of the QAHR when it is biased with current $I$ and the PJVS is biased with frequency $f$. Equation~(\ref{eq.Idirect}) implies that the uncertainty of $I_\mathrm{direct}$ depends in part on the characterization of the QAHR's quantization, as well as that of the Josephson voltage $V_J = N f K_\mathrm{J}^{-1}$. Deviation of $R_{yx}$ from $R_K$ is known to depend on both current bias and electron temperature~\cite{Fox2018,Gotz2018,rodenbach2022,rosen2022,okazaki2022,Lippertz_2022}. 

The currents under test in this work were supplied by the digital current source of a commercial CCC from Magnicon~\cite{gotz2009,drung2009CCC,drung2013CCC}. This source was used to bias the QAHR during current measurements and for in situ characterization of its longitudinal and Hall resistances. The current source features a 16-bit digital-to-analog converter and we used the source's $\pm$300~nA range for all measurements, implying a current resolution of approximately 9~pA. The nominal output of the current source was set using the CCC's graphical user interface (GUI). 

As alluded to in the introduction, we also used the QAHR and an independently-calibrated digital voltmeter (DVM) to perform an in situ calibration of the CCC's current source. We refer to this mode of operation as `indirect' because it relies on the DVM as a transfer standard to complete the measurement itself (as opposed to a transfer standard simply being part of a validation chain). In indirect mode, the PJVS is disabled by disconnecting its bias electronics from the cryostat, and the QAHR functions as a current-to-voltage converter. The circuit arrangement is the same as that shown in Fig.~\ref{fig.1}, except the PJVS voltage is zero and instead of attempting a null, the DVM must be calibrated to accurately measure the relatively small Hall voltage (hundreds of microvolts to several millivolts for nanoampere currents). The calibrated current was then computed using the relationship, 
\begin{equation}
    \label{eq.Iindirect}
    I_\mathrm{indirect} = \frac{V_{DVM}}{R_K},
\end{equation}
where $V_{DVM}$ is the voltage measured by the DVM. We emphasize that the difference between the direct and indirect realizations is not a simple matter of changing voltmeters. In direct mode, radiative leakage from the PJVS microwave bias may couple to other parts of the cryostat, including the QAHR itself (see Section~\ref{section: discussion}). No such concern exists for the indirect realization. 

\section*{3 Realization and validation}\label{section: results}
The first steps after integrating the PJVS and QAHR hardware into the cryostat were to confirm that they functioned as realizations of the volt and the ohm, respectively. For validation of the PJVS, we compared its output voltage against a separately calibrated 8.5 digit DVM. Those results are shown in Table~\ref{PJVS_Budget}. The deviations were all within $\pm2$~$\mu$V/V and within their combined uncertainties. This result is particularly encouraging, considering the rather small voltages and the fact that this check was done after cycling the 1~T field used to magnetize the QAHR. 

\begin{table*}[htb]
\centering
\caption{\textbf{PJVS validation with a calibrated voltmeter.} Shown are the results of a validation of the PJVS with a digital voltmeter (DVM) that had been independently calibrated with a separate PJVS. The first column indicates the number of junctions, $N$, in the subarray used for that row. The second column indicates PJVS bias frequency, $f$. The third column indicates the measured PJVS voltage, $V$. The fourth column is the deviation of the DVM reading from the calculated voltage, $\frac{\delta V}{V}$. The last column shows $U_C$,the combined uncertainty estimate (coverage factor $k=1$) at each voltage.}\label{PJVS_Budget}%
\begin{tabular}{ccccc}
\toprule
$N$ &  $f$ (GHz)  & $V$ (mV)  & $\frac{\delta V}{V} \left(\frac{\mu\textrm{V}}{\textrm{V}}\right)$ & $U_C \left(\frac{\mu\textrm{V}}{\textrm{V}}\right)$ \\
\midrule
12 & 9.701102806 & 0.240723225 	& -0.8 &	27 \\
36 &	9.701102806& 	0.722169675& 	-1.7 &	11 \\
36 &	18.651341000 &	1.388443473 &	1.4 	&6 \\
108& 	9.701102806& 	2.166509025& 	0.4 &	4 \\
108& 	15.300000000 &	3.416888651 &	1.3 &	3 \\ 
108 &	15.302018200 &	3.417339368 &	1.1 &	3\\
324 &	9.701102806& 	6.499527075 &	1.2 &	3 \\
\botrule
\end{tabular}
\end{table*}

\begin{figure}[htb]
\includegraphics[scale=0.9,keepaspectratio]{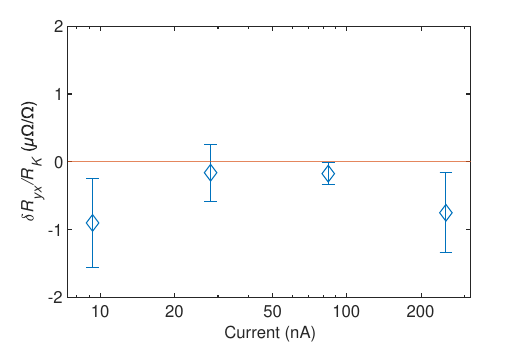}
    \caption{\textbf{Validation of QAHR quantization.} Shown is the relative deviation from quantization, $\delta R_{yx}/R_K = (R_{yx}-R_K)/R_K$ where $R_{yx}$ is the measured Hall resistance of the QAHR, and $R_{K}$ is the von Klitzing constant. The longitudinal bias current for each validation is indicated on the $x$-axis. Each symbol is the average quantization from at least three measurements over a period of six months. In turn, each measurement was comprised of at least 120 observations. The error bars show the standard deviation among the measurements. The combined uncertainty (coverage factor $k=1$) for each symbol is about 1~$\mu\Omega/\Omega$. The PJVS microwave bias was disabled for all four measurements.} 
    \label{fig.Ryx_Validation}
\end{figure}

Validation of the QAHR was established by comparison to a 100~$\Omega$ standard resistor immersed in a temperature-controlled oil bath using bias electronics from Magnicon and a custom CCC helium dip probe. The standard resistor was calibrated by comparison to a GaAs QHRS. Figure~\ref{fig.Ryx_Validation} shows the relative deviation from quantization, $\delta R_{yx}/R_K = (R_{yx}-R_K)/R_K$, of the QAHR for several bias currents over a period of six months. For the data shown, the PJVS's microwave bias was disabled. Though there is a small, negative offset among the set, the average for each current lies within 1~$\mu$V/V of ideal quantization. Additional details of the resistance measurements are presented in the Methods section and in Supplementary Information Sections S2 and S3. 

The third and final part of our evaluation was the quantum realization of the ampere. Six logarithmically-spaced, nominal current values were chosen between 9.3~nA and 252~nA, with an additional five added near the cardinal points for functional testing (see Supplementary Information Section S9). The range of currents was constrained as follows: the maximum bias is limited by current-induced breakdown of the QAH state above some critical value (about 300~nA for the device used in this work). The minimum bias current was based in part on the granularity of the particular PJVS used (i.e., the subarray with the fewest junctions), the CCC's ability to produce an accurate measurement of the QAHR Hall resistance at that current, and on signal-to-noise limitations (e.g., the DVM used for validation of the PJVS and for indirect current measurements). As suggested by Equation~\ref{eq.Iindirect}, our computation of $I_\mathrm{indirect}$ did not correct for the small QAHR quantization offsets shown in Fig.~\ref{fig.Ryx_Validation}. Instead, we used the defined constants of the SI for the denominator $R_K$. Though the indirect method does share the same QAHR as its direct counterpart, each is a valid, practical realization of the ampere~\cite{ampere_miseenpratique}. Therefore, any disagreement between the direct and indirect ampere realizations is indicative of the metrological performance of our prototype. 

\begin{figure*}[h]
    \centering    \includegraphics[width=\textwidth,keepaspectratio]{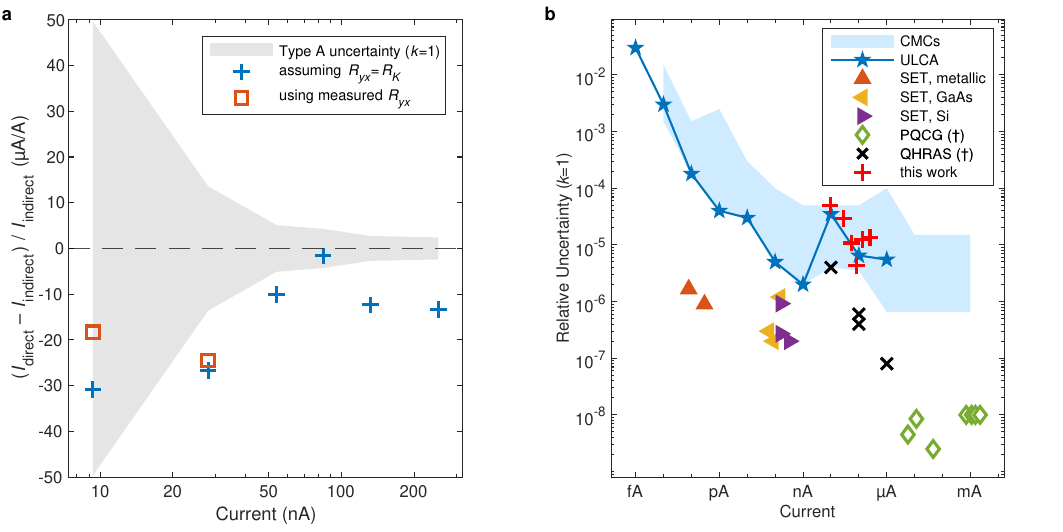}
    \caption{\textbf{Direct realization of the ampere and contextualization.} \textbf{(a)} Relative disagreement between $I_\mathrm{direct}$ and $I_\mathrm{indirect}$: $\left(I_\mathrm{direct}-I_\mathrm{indirect}\right)/I_\mathrm{indirect}$. Disagreement shown is the weighted mean of multiple measurements. Blue crosses show results assuming ideal quantization of the QAHR (i.e., $R_{yx} = R_K$). Orange squares (for the two lowest currents) show disagreement when using a value for $R_{yx}$ determined from separate Hall resistance measurements performed while the PJVS microwave bias was active. Shaded area shows the combined type A (statistical) uncertainty of the direct and indirect measurements. \textbf{(b)} Worldwide comparison of relative uncertainties reported for precision current generation and measurement within the range ($10^{-15}$~A to $10^{-3}$~A). The shaded region shows calibration and measurement capabilities (CMCs) using Ohm's law or capacitive charging methods~\cite{BIPM2023} reported by nine national metrology institutes (see main text). Also shown are CMCs from PTB and INRIM based on a calibrated ULCA~\cite{BIPM2023} (blue stars with a solid blue line as a guide to the eye). Uncertainties for single electron transport (SET) realizations are shown for metallic~\cite{SETmet1,SETmet2} (green triangles), GaAs~\cite{SETGaAs1,SETGaAs2,SETGaAs3,Stein2017} (orange triangles), and silicon~\cite{SETSi1,SETSi2,SETSi3} (purple triangles) devices. Laboratory-scale integrated PJVS-QHR ampere realizations are denoted in the legend by the $\dagger$ symbol and include the PQCG (green diamonds)~\cite{Brun-Picard_2016,Djordjevic_2025} and realizations based on quantum Hall resistance array standards (QHRAS, black x symbols)~\cite{Chae_2020,Chae_2022}. Combined uncertainties for our single-cryostat ampere realization are shown as red crosses. }
    \label{fig.ampere_comp1}
\end{figure*}

Figure~\ref{fig.ampere_comp1}a shows the relative disagreement, $\left(I_{\mathrm{direct}} -I_{\mathrm{indirect}}\right)/I_{\mathrm{indirect}}$, of the two modes of current measurement. Currents within 500~$\mu$A$/$A of one another were grouped by their nominal values to aid visualization. A table showing the exact GUI setting for $I$, and the corresponding, individually-measured values of $I_{\mathrm{direct}}$ and $I_{\mathrm{indirect}}$, can be found in Section S7 of the Supplementary Information. The blue crosses depict the disagreement assuming that $R_{yx} = R_K$ for \emph{both} the direct and indirect measurements. For the two lowest currents, we also show (orange squares) the disagreement if we substitute measured values for $R_{yx}$ in the computation of $I_\mathrm{direct}$, according to Equation~\ref{eq.Idirect} (the computation of $I_\mathrm{indirect}$ remains that of Equation~\ref{eq.Iindirect}). For the two points shown, $R_{yx}$ was measured by enabling the PJVS microwave bias---but keeping its output voltage at zero---and performing a measurement of the QAHR's Hall resistance with the CCC. 

For this intercomparison, we performed a statistical, or Type A, evaluation of uncertainty~\cite{GUM} by computing the standard error for each realization of the ampere at each current and combining them by propagating uncertainties through the relative disagreement relation. The bounds for a coverage factor $k=1$ (corresponding to one standard deviation of the underlying distribution) is shown as a shaded region centered around 0~$\mu$A$/$A. For most of the currents evaluated, the observed disagreements between the two modes lie outside the Type A bounds, suggesting an unresolved systematic error. 

We attribute this error to imperfect electromagnetic shielding of radiation from the PJVS's coplanar waveguide. In preliminary measurements, we observed that many hours were required before the QAHR's Hall resistance approached an equilibrium value either during or after microwave exposures (see following section). Interestingly, substituting measured values for $R_{yx}$ for each of the two lowest currents, according to Equation~\ref{eq.Idirect}, reduced the disagreement. In other words, the correction had the proper sign, but did not completely eliminate the disagreement. Due to the time constraints of sharing the cryostat with other users, we could not wait for full thermalization at each current. Minimizing radiative microwave leakage and improving thermalization will be a focus of future work. 

Fig.~\ref{fig.ampere_comp1}b contextualizes our unified realization of the ampere against published uncertainties worldwide. National metrology institutes (NMIs) report their regularly achievable calibration and measurement capabilities (CMCs) to a database maintained by the Bureau International des Poids et Mesures (BIPM)~\cite{BIPM2023}. Shown are CMCs (blue shaded region) from nine NMIs for a broad range of currents (the full list of NMIs can be found in Supplementary Information Section S1). State-of-the-art CMCs from the Physikalisch-Technische Bundesanstalt (PTB) and the Istituto Nazionale di Ricerca Metrologica (INRIM) using an ultrastable low-noise current amplifier (ULCA)~\cite{BIPM2023} are depicted as blue stars. We also show uncertainties for several quantum realizations of the ampere, including those based on single-electron transport (SET) devices~\cite{SETmet1,SETmet2,SETGaAs1,SETGaAs2,SETGaAs3,Stein2017,SETSi1,SETSi2,SETSi3} (triangles) and those based on laboratory-scale integration of a PJVS and a QHRS~\cite{Brun-Picard_2016,Djordjevic_2025} (green diamonds) or QHRAS~\cite{Chae_2020,Chae_2022} (black x symbols). The relative combined uncertainties of our single-cryostat realization of the ampere (direct mode, red crosses) are comparable to state-of-the-art CMCs for all currents evaluated. 

\section*{4 Limitations of the direct realization of the ampere}\label{section: discussion}
In our prototype, we observed that the PJVS microwave bias could lead to increases in the QAHR's longitudinal resistivity $\rho_{xx}$, the latter often considered a proxy for the electron temperature within the QAHR heterostructure~\cite{Bestwick2015}. Though the PJVS cryopackage is highly engineered, it is common for such devices to exhibit frequency-dependent reflections at the microwave input. Even if the reflection coefficient is comparatively low, a small fraction of incident microwave power may be radiated and couple to other stages of the cryostat (e.g., through wiring or apertures). Similar effects have been reported to occur when integrating a PJVS into a cryogenic electrical substitution radiometer~\cite{White_2024}. But the detector in that system is operated at much higher temperatures than our QAHR. By shielding the PJVS in our unified realization and restricting the set of bias frequencies, the impact of radiative leakage was minimized, but could not be eliminated altogether (see Supplementary Information Section S11 for details). If such leakage diminished quantization of the QAHR during direct mode, it could explain the disagreement between the two realizations of the ampere. 

\begin{figure*}[htb]
    \centering    \includegraphics[width=\textwidth,keepaspectratio]{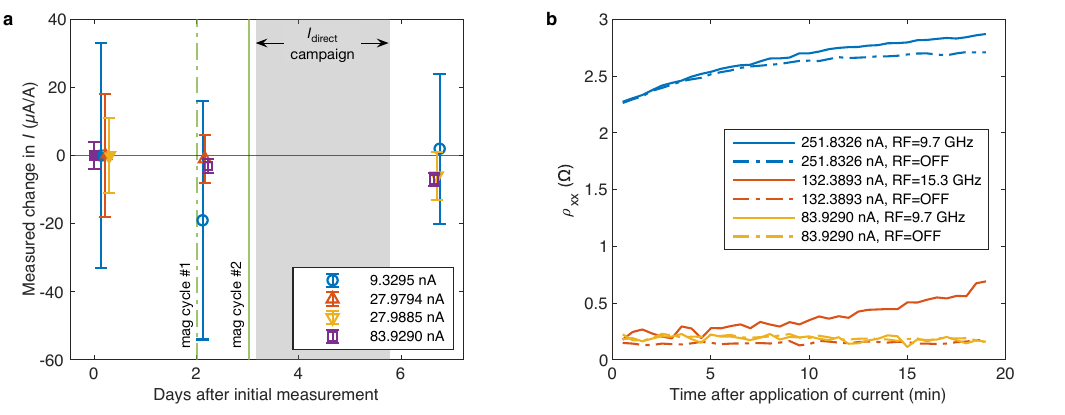}
    \caption{\textbf{Time dependence of indirect realizations of the ampere and of QAHR longitudinal resistivity $\rho_{xx}$}. \textbf{(a)} Measured change in current $I$ for different measurement methods and experimental conditions. Initial measurements of the current source (filled symbols, around day 0) were performed by directly connecting the calibrated DVM and using its DC current function (100~nA range), thus bypassing the QAHR. These symbols lie on the x-axis since they are the reference values for subsequent measurements (open symbols) of the current source using the indirect method (i.e., via QAHR and DVM's voltage function). Each symbol represents an average of at least 15 observations. Error bars indicate the Type A uncertainty for each measurement of current. The gray shaded region depicts the period over which all direct ampere realizations were performed. QAHR magnetization cycles on day 2 and day 3 (shown by green vertical lines) preceded the direct realization campaign. Indirect realizations on either side of the direct measurement period show repeatability within $\pm 5$~$\mu$A/A for 27.9885~nA and for 83.9290~nA. For 9.3295~nA, indirect estimates lie within $\pm 10$~$\mu$A/A over the same span of time. Agreement was quantitatively similar between the initial DVM current function measurements and indirect realizations after the first magnetization cycle. \textbf{(b)} Change in QAHR longitudinal resistivity $\rho_{xx}$ after application of current bias under varying microwave bias (indicated by ``RF'' in the legend). Data were acquired on day 6, with most recent magnetization occurring on day 3. The longitudinal voltage was measured with the same nanovoltmeter used in the realization of $I_\mathrm{direct}$. For the case of zero microwave bias, all currents $\ge$83.9290~nA showed a measurable $\rho_{xx} > 0$. }
     \label{fig.IndirectRxxDrift}
\end{figure*}

If direct and indirect measurements of the same current setpoint were separated in time, it is also possible that the disagreement in the comparison is just due to drift in the output from the CCC's current source. Fig.~\ref{fig.IndirectRxxDrift}a shows the relative change in $I$ for different measurement techniques and experimental conditions. The four filled symbols near day 0 show measurements of the current source by direct connection to a calibrated DVM and using its DC current function (i.e., bypassing the QAHR). Open symbols show the relative difference between $I_\mathrm{indirect}$ and those initial, day 0 measurements. Also shown (green vertical lines) are QAHR magnetization cycles and the period over which direct realizations were performed and thus when microwave leakage was present (gray shaded region). 

The relative change in the measured current over the entire period shown in Fig.~\ref{fig.IndirectRxxDrift}a is less than $\pm$5~$\mu$A/A for 27.9885~nA and for 83.9290~nA, and approximately $\pm$10~$\mu$A/A for 9.3295~nA. The larger relative change at 9.3295~nA may have resulted from the much larger statistical fluctuations in the measurements (shown as error bars and based on at leats 15 observations). For the two lowest currents, the relative change in $I_\mathrm{indirect}$ is less than half the disagreement between $I_\mathrm{direct}$ and $I_\mathrm{indirect}$. For 83.9290~nA, the relative standard deviation of the points shown in Fig.~\ref{fig.IndirectRxxDrift}a is $\sigma_I/I = 4$~$\mu$A/A. Though this is about twice the disagreement, it is comparable to the statistical uncertainty of the corresponding direct realization ($3.8$~$\mu$A/A). 

The stability of $I_\mathrm{indirect}$ shown in Fig.~\ref{fig.IndirectRxxDrift}a allows us to draw two conclusions about our prototype. First, current source drift alone cannot account for the magnitude of the disagreement in our two realizations of the ampere. Further, the disagreement was uniformly negative, which is readily understandable if the result of microwave leakage is a reduced $R_{yx} < R_K$. Second, since we assumed $R_{yx} = R_K$ for all computations of $I_\mathrm{indirect}$, the observed stability means we can rule out a permanent, microwave-induced loss of quantization for the QAHR, at least for the main device under consideration. However, we cannot rule out ephemeral changes to the QAHR's quantization during direct realizations. 

Fig.~\ref{fig.IndirectRxxDrift}b depicts measurements of $\rho_{xx}$ performed after the direct campaign shown in Fig.~\ref{fig.IndirectRxxDrift}a, with and without microwave exposure. The time evolution of $\rho_{xx}$ in the presence of microwave leakage implies a similar effect on $R_{yx}$ via the relation $\delta R_{yx} = -s\rho_{xx}$, where $s$ is the dimensionless admixing parameter (typically between 0.01 and 1)~\cite{okazaki2022,Patel2024}. Only at the highest bias current of 251.8~nA do we find notable, current-induced dissipation (RF OFF). Repeating that measurement with the microwave bias at the optimized frequency of 9.7~GHz, we see a slight increase in $\rho_{xx}$ compared to its baseline. Conversely, at $I = 132.4$~nA and the less optimal $f=15.3$~GHz, we find a more noticeable rate of increase in dissipation. Despite those two currents showing comparable disagreements (Fig.~\ref{fig.ampere_comp1}a), we note that they are probably due to differing effects. At 251.8~nA and 9.7~GHz, quantization is likely diminished due to the combined effects of current- and microwave-induced breakdown. But at 132~nA and 15.3~GHz, it is likely that microwave leakage is the dominant source of breakdown. 

Though microwave leakage limited the accuracy of our initial prototype, we find the achievement of a 4.3~$\mu$A/A uncertainty at 83.9~nA promising. Our efforts highlight the substantial engineering challenge of co-locating such quantum standards in a single cryostat. We expect future iterations with dedicated cryogenics and improved shielding will no doubt eliminate the observed systematic effects. 

\section*{5 Evaluation of uncertainty for the ampere}\label{section: uncertainty}
The overall uncertainty budget for our direct realization of the ampere is given in Table~\ref{tab. direct budget}. Each component is comprised of one or two types of uncertainty evaluation: Type A (statistical) or Type B (other than statistical). We define our Type A uncertainty as the standard error of the null voltage readings from a single experimental run. The Type A uncertainty decreases as the magnitude of the bias current increases because we are limited by the voltage noise floor of our prototype. Allan deviation measurements (Section S12 of the Supplementary Information) suggest that the noise spectrum was not completely white and therefore additional averaging may not have reduced our Type A uncertainty. Those measurements also suggest that the long cabling between the cryostat and nanovoltmeter may be a source of excess noise. 

Type B evaluations include estimates for measurement of $\delta R_{yx}$, microwave leakage effects, offsets to measurements of $V_{\mathrm{null}}$ due to the finite input impedance, gain and linearity of the nanovoltmeter, and uncertainty of the PJVS microwave bias frequency $f$. The combined root-sum-squared (RSS) uncertainty for each nominal current is given in the final row of the table. All uncertainties listed are for a coverage factor $k = 1$. A detailed description of each uncertainty component can be found in Section S4 of the Supplementary Information. 

\begin{table*}[htb]\centering
\caption{\textbf{Uncertainty budget for measurements of $I_{\mathrm{direct}}$.} Columns one and two indicate the uncertainty component and the type of evaluation(s) performed, respectively. Column three shows the relative contribution (in units of $\mu$A/A) for each nominal current. The contribution for `$R_{yx}$ quantization' includes estimates for the effects of microwave leakage on quantization of the QAHR. Listed uncertainties  assume a coverage factor $k=1$. }\label{tab. direct budget}%
\begin{tabular}{@{}lcccccccccc@{}}
\toprule
Component &  Type  && \multicolumn{6}{c}{Contribution $\left(\frac{\mu\textrm{A}}{\textrm{A}}\right)$} \\
\midrule
Nominal current (nA) & --- && 9.3 & 28.0 & 53.8 & 83.9 & 132.4 & 251.8 \\
\midrule
Null detector readings & A && 39 & 12 & 3.9 & 3.8 & 2.2 & 2.3 \\
$R_{yx}$ quantization & {A+B} && 31 & 27 & 10 & 1.9 & 12 & 13 \\
Nanovoltmeter input impedance & B && \multicolumn{6}{c}{0.005} \\
Nanovoltmeter gain \& linearity & B && \multicolumn{6}{c}{0.03} \\
PJVS microwave bias frequency & B && \multicolumn{6}{c}{0.01}\\
\midrule
Total & RSS && 50 & 29 & 11 & 4.3 & 13 & 14 \\
\botrule
\end{tabular}
\end{table*}

\section*{6 Conclusions}\label{section: conclusion}
We have successfully demonstrated quantum realizations of the SI volt, ohm, and ampere from a single cryostat containing a PJVS and a QAHR. The unified instrument exhibited uncertainties comparable to NIST’s CMCs for low voltages (2~$\mu$V/V) and to recent precision measurements of the ohm on other QAHR devices (about 1~$\mu\Omega/\Omega$). For direct realizations of the ampere, our best uncertainty was 4.3~$\mu$A/A at 83.9~nA. Increased uncertainties at low (50~$\mu$A/A at 9.3~nA) and high (14~$\mu$A/A at 252~nA) currents was attributed to microwave leakage from the PJVS. Even so, our uncertainties were comparable to well-established measurement capabilities at several national metrology institutes. 

Our next efforts will be focused on improvements to the cryostat and a comparison against an independently-calibrated current source. Optimized electromagnetic shielding and filtering will eliminate the effect of microwave leakage on the QAHR. Improved wiring inside and outside the cryostat will reduce the effects of triboelectric and thermoelectric noise, which should ultimately reduce statistical uncertainty. Likewise, designing and fabricating wider Hall bar structures should increase the current capacity into the microampere regime. Finally, improved device thermalization and continued efforts to understand current breakdown mechanisms will ensure a more robust topological state in the QAHR. 

Our hope is that this initial effort towards a multifunction quantum instrument will spur innovation in topological material systems and in cryostat design and engineering. With continued advancement, a more robust instrument may be possible, capable of operation above 1~K and at zero magnetic field, thereby greatly reducing the infrastructure required to disseminate electrical quantities from the quantum SI.  

\section*{7 Methods}\label{section:Methods}

\noindent\textbf{PJVS and Voltmeter Characterization} 
The PJVS device used in this work was based on the same triple-stacked, Nb/Nb$_x$Si$_{1-x}$/Nb junction process used in the National Institute of Standards and Technology's (NIST) Standard Reference Instruments~\cite{Afox2014}. The cryopackaged chip contains 59,140 junctions and is capable of an output voltage of approximately 2~V at a microwave bias frequency of 18~GHz. The junctions are arranged into two symmetric arrays and share a common microwave bias, which is distributed using on-chip microwave splitters. This allows the chip to be operated as a dual-channel (1~V + 1~V) source and was the configuration used in this work. Each array is further divided into ternary-weighted subarrays. 

The magnetic field needed to polarize the QAHR into its quantized state raises concerns regarding the operation of the PJVS. Josephson junction arrays can suffer diminished critical currents and/or operating margins when subjected to magnetic fields~\cite{Afox2019}. Though resistance measurements of the QAHR and indirect-mode current were performed in zero magnetic field, it is possible that the initial magnetization process could result in trapped flux in the PJVS or otherwise impair its operation. 

Trapped flux can be resolved by heating the Josephson junctions above their critical temperature (about 9~K for the junctions used in this work). However, heating the 4~K stage of the dilution refrigerator (DR) could produce a high mixing chamber (MXC) temperature, possibly demagnetizing the QAHR~\cite{rodenbach2021}. It would therefore be impossible to achieve simultaneous quantization of the PJVS and QAHR.  We checked for this possibility by elevating the 4~K stage to 12~K, which required briefly disabling the pulse tube cryocooler. The procedure was successful with only minimal heating of the MXC ($T$ remained below 20~mK). Upon further evaluation (see Section S5 of the Supplementary Information), we determined that the impact of the magnetizing field on the PJVS was minimal and reversible, and thus such temperature cycling was rarely required (e.g., after a temporary power failure). 

We used a commercial nanovoltmeter (Keithley 2182A) to measure the null voltage during direct realization of the ampere (i.e., measurements of $I_\mathrm{direct}$). The nanovoltmeter was calibrated using NIST's conventional digital multimeter calibration service to provide a bound on the meter's performance. As outlined in S4 of the Supplementary Information (see uncertainty budget discussion), the corrections are negligible for the purposes of our overall budget for $I_\mathrm{direct}$. Additional details concerning the meter's Allan deviation performance can also be found in Section S12 of the Supplementary Information. 

We also used a separately-calibrated 8.5-digit digital voltmeter, or DVM (Keysight 3458A) to (a) validate the PJVS within the DR and (b) measure the transverse Hall voltage $V_{yx}$ of the QAHR during measurements of $I_{\mathrm{indirect}}$. As with the nanovoltmeter, we did not apply corrections to the DVM's readings. After the DVM's initial calibration, we periodically checked its performance against the PJVS within the DR to ensure that disagreements between the calculated PJVS voltage and the DVM reading were within an acceptable range---less than the measurement's expanded Type A uncertainty. This was facilitated by a separate set of voltage leads that run from the cryopackage directly to room-temperature connections (i.e., $R_{yx}$ is not in the measurement loop). \\

\noindent\textbf{QAH film composition and device fabrication}The Cr-doped (Bi,Sb)$_2$Te$_3$ (Cr-BST) material used in this work was grown on an epi-ready semi-insulating GaAs (111)B substrate in an ultra-high vacuum molecular beam epitaxy (MBE) system. Before growth, the substrate was loaded into the MBE chamber and pre-annealed at a temperature of 630 \textdegree C in a Te-rich environment to remove the oxide on the surface. During growth, the substrate was kept at 200 \textdegree C and in-situ reflection high-energy electron diffraction (RHEED) was used to monitor the quality and thickness of the film. The Cr-BST film consisted of six quintuple layers (QLs): The first and sixth QLs had increased Cr concentration, (Cr$_{0.24}$Bi$_{0.26}$Sb$_{0.62}$)$_2$Te$_3$, compared to the inner four QLs, (Cr$_{0.12}$Bi$_{0.26}$Sb$_{0.62}$)$_2$Te$_3$. 

The QAHR devices used in this work were fabricated using the following steps: (1) chip cleaned in acetone and isopropyl alcohol, (2) SPR 3612 photoresist spun at 5.5~krpm for 45~s and baked at 80~°C for 2~minutes, (3) resist removed using phosphate salt developer and two water rinses, (4) 5~nm / 90~nm Ti/Au metals deposited using electron-beam (e-beam) evaporation following a 10~s in situ Ar pre-etch, (5) metal lift-off in acetone, (6) 1~nm Al seed layer deposited at 0.02~nm/s, (7) 40~nm alumina gate dielectric deposited via low temperature atomic layer deposition, (8) gate electrode patterned via photolithography, (9) gate metals (5~nm / 80~nm Ti/Au) deposited via e-beam evaporation, (10) final metal liftoff, and (11) masked wet alumina etch for 150~s (mask defined via photolithography and tetramethylammonium hydroxide (TMAH)-based developer). \\

\noindent\textbf{Precision measurements of QAHR.} Precision metrological characterization of the QAHR was performed with a commercial cryogenic current comparator (CCC) bridge (Magnicon GmbH, see e.g.,~\cite{gotz2009}). Additional details of the methodology used can be found in earlier works~\cite{Fox2018,rodenbach2022}. Prior to characterization of its longitudinal and Hall resistivities ($\rho_{xx}$ and $\rho_{yx}$, respectively), the QAHR was polarized by applying a (0.5 to 1)~T out-of-plane magnetic field and then reducing the field to 0~T. A typical measurement sequence consisted of at least 30 polarity reversal cycles (to minimize thermoelectric voltages and offset drift). For each polarity, 60 individual measurement points were collected, the first 14 of which are discarded to account for settling after reversal. All reported measurements were performed at the cryostat's base temperature of approximately 10~mK. 

Measurements of $\rho_{xx}$ were performed with the QAHR in the CCC's first current loop, $I_1$, and the second current loop, $I_2$, shorted. The CCC’s compensation network and the digital current source for $I_2$ were both inactive. The QAHR was then biased using the current source for $I_1$ and its longitudinal voltage measured by the CCC's built-in nanovoltmeter. The dependence of $\rho_{xx}$ on top gate voltage, $V_\mathrm{gate}$, was determined by biasing the QAHR at $I_1=100$~nA and sweeping $V_\mathrm{gate}$. The gate sweep allowed us to determine the native Fermi level of our heterostructure. As shown in Extended Data Fig.~1, $\rho_{xx}$ exhibited a local minimum near $V_\mathrm{gate} = 0$~V. Therefore, all realizations of the SI ampere and ohm were performed with $V_\mathrm{gate}=0$~V (i.e., gate held at ground). Extended Data Fig.~1 also shows that $\rho_{xx}$ gradually decreased over the course of several hours after a magnetization cycle. This may be an indication of slow thermal relaxation processes within the heterostructure. 

To evaluate QAHR quantization, we operated the CCC in bridge mode, comparing the QAHR's transverse resistance ($R_{yx}$) to a calibrated and temperature-regulated 100~$\Omega$ standard resistor. From the known current ratio in the two arms of the bridge (enforced by the CCC), and the measured difference between the QAHR's Hall voltage and the voltage across the standard resistance, the quantization of the QAHR relative to $R_K$, $\delta R_{yx} = (R_{yx} - R_K)/R_K$, may be computed. 

To assess current-induced breakdown, we measured $\rho_{xx}$ and $\delta R_{yx}$ at bias currents ranging from 9~nA to 252~nA. We repeated some of these measurements with the PJVS microwave bias enabled (i.e., $f \neq 0$ but $V_J$ = 0 V), with the goal of quantifying microwave-induced breakdown on the QAHR. Extended Data Fig.~2 illustrates the potential impact of different microwave bias frequencies on QAHR quantization. Further details on current- and microwave-induced breakdown can be found in Supplementary Information Section S11. \\

\noindent\textbf{Direct realization of the ampere} 
Direct realization of the ampere $I$ was based on comparing the QAHR's Hall voltage $V_{yx} = I R_{yx}$ against a nominally similar voltage from the PJVS $V_J$, and measuring the residual null voltage with a nanovoltmeter. To eliminate thermoelectric offsets in $V_{\mathrm{null}}$ the polarity of $I$ was reversed periodically (typically every 60~s) for a total of $n$ cycles ($10 \leq n \leq 90$). In each cycle, we delayed acquisition for 10~s to allow for settling and then measured approximately 40 individual voltage points. Extended Data Fig.~3 shows an example measurement of $V_{\mathrm{null}}$ for $I = 83.9$~nA. 

$V_{\mathrm{null}}$ used in Equation~(\ref{eq.Idirect}) of the main text was determined as follows: First, two averages are computed using the raw null voltage readings $V^\mathrm{raw}_{\mathrm{null}}$ (show as red dots in Extended Data Fig.~3a) measured during the polarity reversal process. These averages are $V^+_\mathrm{null}$ and $V^-_\mathrm{null}$. $V^+_\mathrm{null}$ ($V^-_\mathrm{null}$) is the average value of $V^\mathrm{raw}_{\mathrm{null}}$ for a positive (negative) portion of a polarity reversal cycle (shown as blue (black) triangular symbols in Extended Data Fig.~3a). The difference between $V^+_\mathrm{null}$ and $V^-_\mathrm{null}$ for consecutive half-cycles is then,
\begin{equation}
\label{eq:Vdiff}
V_{\mathrm{diff}}= \frac{1}{2}\left(V_{\mathrm{null}}^+ -V_{\mathrm{null}}^-\right). 
\end{equation}
These $2n-1$ measurements of $V_\mathrm{diff}$ are again averaged using an inverse variance weighting to produce a final value, $V_\mathrm{null}$. It is this value, $V_\mathrm{null}$, which is used to calculate $I_\mathrm{direct}$ via Equation~(\ref{eq.Idirect}). Extended Data Fig.~3 and its caption illustrate the procedure. 

\section*{8 Extended Data}
\setcounter{figure}{0}

\renewcommand{\thefigure}{E\arabic{figure}}
\begin{figure}[H]
    \begin{centering}  \includegraphics[scale=0.8,keepaspectratio]{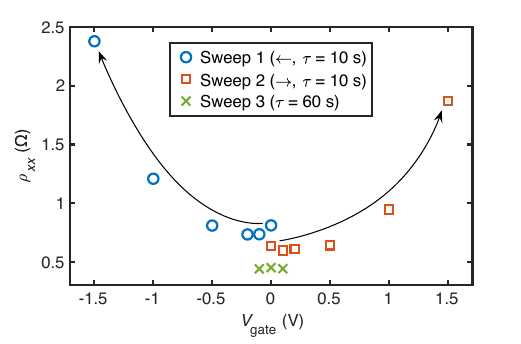}
        \caption{\textbf{Longitudinal resistivity as a function of top gate voltage.} Shown are measurements of QAHR longitudinal resistivity, $\rho_{xx}$, versus top gate voltage $V_\mathrm{gate}$, performed with the CCC's nanovoltmeter. First, $V_\mathrm{gate}$ was quickly swept toward negative voltages (blue circles, integration time $\tau=10$~s). $V_\mathrm{gate}$ was next brought back to zero and then swept toward positive voltages (orange squares, $\tau=10$~s). Each symbol represents an average of 30 observations per measurement. The time between measurements (symbols) was approximately 7 minutes. The curved arrows are intended as guides to the eye and show the sweep direction. Finally, $V_\mathrm{gate}$ was slowly swept over a narrow range (green crosses, $\tau=60$~s). For the narrow sweep, each symbol represents an average of 10 observations and the time between measurements was 12 minutes. The local minimum in $\rho_{xx}$ seen near $V_\mathrm{gate}=0$~V indicates that the native Fermi level is well positioned near the center of the magnetic exchange gap. The mixing chamber temperature over the course of these sweeps (lasting about two hours total) only decreased by 0.2~mK. The decrease in $\rho_{xx}$ near $V_\mathrm{gate} = 0$ with each sweep suggests that either $\rho_{xx}$ is strongly sensitive to temperature or that thermalization occurs over very long time scales.}
        \label{fig.rxx_v_vg}
    \end{centering}
\end{figure} 

\begin{figure}[H]
    \begin{centering}  \includegraphics[scale=0.75,keepaspectratio]{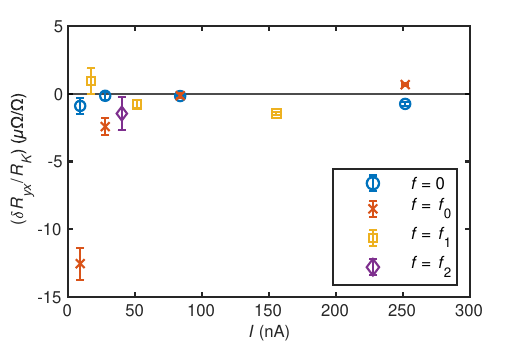}
        \caption{\textbf{Characterization of QAHR quantization in the presence of microwave leakage.} Deviation of the Hall resistance from quantization, $\delta R_{yx}/R_K = (R_{yx}-R_K)/R_K$, was measured as a function of bias current $I$ and PJVS microwave bias. The impact of the microwave leakage varied with the microwave frequency $f$.  For $f=0$ (blue circles, microwave bias disabled), each symbol represents the average of at least three measurements, each based on at least 30 observations. For other frequencies only one measurement (again, based on at least 30 observations) was performed for each current. Error bars show the Type A (statistical) uncertainty. Note that the uncertainty for some points may be smaller than the symbol. For the $f\neq0$ measurements, the PJVS microwave excitation was active (input power $P=1$~mW), but the PJVS output voltage was zero. The frequencies $f_n$ in the legend are as follows: $f_0 = 9.701$~GHz (orange crosses), $f_1 = 18.00$~GHz (yellow squares), and $f_2 = 14.00$~GHz (violet diamonds).}
        \label{fig.ryx CCC}
    \end{centering}
\end{figure} 

\begin{figure}[H]
    \begin{centering}  \includegraphics[scale=0.65,keepaspectratio]{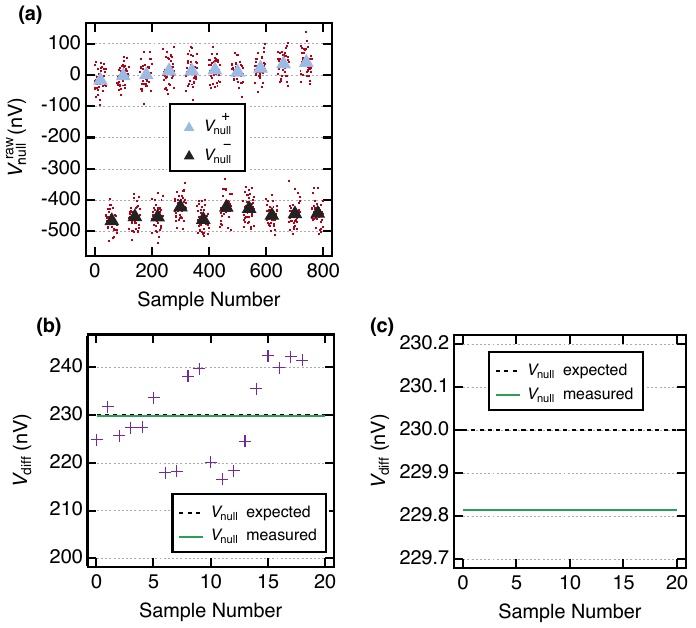}
        \caption{\textbf{Example raw measurement data of $V_\mathrm{null}$.} Shown are raw $V_\mathrm{null}$ data taken for $I = 83.9$~nA. \textbf{(a)} 800 Individual measurements of $V^\mathrm{raw}_\mathrm{null}$ over several polarity reversal cycles are shown as red dots. Each raw measurement was acquired using a 60~s integration time. The triangular symbols show $V^+_\mathrm{null}$ ($V^-_\mathrm{null}$), the average value of the raw null voltage readings for the positive (negative) portion of a polarity reversal cycle. \textbf{(b)} The 18 measurements of $V_\mathrm{diff}$ calculated from the data in \textbf{(a)} via Equation~(\ref{eq:Vdiff}) are shown as cross-symbols. The solid green line is $V_\mathrm{null}$---the average of the 18  $V_\mathrm{diff}$ values. The dashed line shows the corresponding null voltage expected from the indirect realization of the ampere, $V_\mathrm{null,\,expected} = I_\mathrm{indirect}R_{K} - NfK_{J}^{-1}$. \textbf{(c)} An expanded view of the two trendlines shown in \textbf{(b)}}.
        \label{fig.vnull}
    \end{centering}
\end{figure} 
\section*{9 Data Availability}
Data that support the findings of this study are included as supplementary information for this article. Raw data files (e.g., CCC output files) are available from the corresponding authors upon reasonable request.

\backmatter
\bmhead{Supplementary Information}
See supplementary material for additional details, such as the names of the national metrology institutes referenced for calibration and measurement capability data, the indirect current measurement schematic, a discussion of the cryogenic current comparator's digital current source, tables detailing how measurements of $I_{\mathrm{direct}}$, $I_{\mathrm{indirect}}$ correspond to the GUI setting of $I$, evidence of radiative heating of the QAHR via microwave leakage, and Allan deviation calculations.\\

\bmhead{Acknowledgments}
The authors thank T. Mai, F. Fei, G. J. Fitzpatrick, and E. C. Benck for assistance with the NIST internal review process. The authors also thank Ilan T. Rosen and Marc A. Kastner for enlightening discussions throughout this work. L.K.R., M.P.A. and D.G.-G.  were supported by the Air Force Office of Scientific Research (AFOSR) Multidisciplinary Research Program of the University Research Initiative (MURI) under grant number FA9550-21-1-0429. At the initiation of the project L.K.R., M.P.A. and D.G.-G. were supported by the U.S. Department of Energy, Office of Science, Basic Energy Sciences, Materials Sciences and Engineering Division,under Contract No. DE-AC02-76SF00515 and the Gordon and Betty Moore Foundation through Grant No. GBMF9460. P.Z., L.T. and K.L.W. acknowledge the support from the National Science Foundation (NSF) under the Accelerating Interdisciplinary Frontiers in Quantum Sciences and Technologies (grant 2125924, NRT-QISE) and Quantum Devices with Majorana Fermions in High-Quality Three-Dimensional Topological Insulator Heterostructure (grant 1936383, QII-TAQS) programs. P.Z., L.T. and K.L.W. were also supported by the Army Research Office MURI under grant numbers W911NF16-1-0472 and W911NF-19-S-0008. 
Commercial equipment, instruments, and materials are identified in this paper in order to specify the experimental procedure adequately. Such identification is not intended to imply recommendation or endorsement by the National Institute of Standards and Technology or the United States government, nor is it intended to imply that the materials or equipment identified are necessarily the best available for the purpose. Work presented herein was performed, for a subset of the authors, as part of their official duties for the United States government. Funding is hence appropriated by the United States Congress directly. Part of this work was performed at nano@stanford, supported by the National Science Foundation under award ECCS-2026822.

\bmhead{Author Contributions Statement} 
L.K.R., J.M.U., D.B.N., A.F.R., and D.G.-G. conceived and designed the experiments. L.K.R., J.M.U., N.T.M.T., A.R.P., M.P.A., and Z.S.B. performed the experiments. L.K.R., J.M.U., N.T.M.T., A.R.P., Z.S.B., and D.G.-G., analyzed the data. P.Z., L.T., and K.L.W. contributed MTI thin film materials. L.K.R. and M.P.A. fabricated and qualified QAH devices. L.K.R., J.M.U., N.T.M.T., A.R.P., S.U.P., D.G.J, R.E.E., D.B.N., A.F.R., and D.G.-G. contributed specialized hardware and expertise to support metrology experiments. L.K.R. and J.M.U. wrote the paper. N.T.M.T., A.R.P., M.P.A., Z.S.B., A.F.R., D.G.-G. reviewed and provided input on the paper. 

\bmhead{Competing Interests Statement}
The authors declare no competing interests.

\providecommand{\noopsort}[1]{}\providecommand{\singleletter}[1]{#1}%


\newpage
\onecolumn
 \begin{flushleft}
        \Large\textbf{Supplemental Information for: A unified realization of electrical quantities from the quantum International System of Units}
    \end{flushleft}
    \vspace{2ex}
    \begin{flushleft}
        {Linsey K. Rodenbach$^{1,2\dag}$, Jason M. Underwood$^{3\dag *}$, Ngoc Thanh Mai Tran$^{3,4}$, Alireza R. Panna$^3$, Molly P. Andersen$^{2,5}$, Zachary S. Barcikowski$^{3,4}$, Shamith U. Payagala$^3$, Peng Zhang$^6$, Lixuan Tai$^6$, Kang L. Wang$^6$, Dean G. Jarrett$^3$, Randolph E. Elmquist$^3$, David B. Newell$^3$, Albert F. Rigosi$^3$ and David Goldhaber-Gordon$^{1,2*}$}
    \end{flushleft}
    \vspace{2ex}
    \begin{flushleft}
        \footnotesize{$^1$\textit{Department of Physics, Stanford University, Stanford, California 94305, USA}}\\
        \footnotesize{$^2$\textit{Stanford Institute for Materials and Energy Sciences, SLAC National Accelerator Laboratory, 2575 Sand Hill Road, Menlo Park, California 94025, USA}}\\
        \footnotesize{$^3$\textit{Physical Measurement Laboratory, National Institute of Standards and Technology (NIST), Gaithersburg, Maryland 20899, USA}}\\
        \footnotesize{$^4$\textit{Joint Quantum Institute, University of Maryland, College Park, Maryland 20742, USA}}\\
        \footnotesize{$^5$\textit{Department of Materials Science and Engineering, Stanford University, Stanford, California 94305, USA}}\\
        \footnotesize{$^6$\textit{Department of Electrical and Computer Engineering, University of California, Los Angeles, Los Angeles, California 90095, USA}}\\       
        \footnotesize{$^\dag$These authors contributed equally to this work.}\\
        \footnotesize{$^*$To whom correspondence should be addressed; E-mail: \texttt{jason.underwood@nist.gov and goldhaber-gordon@stanford.edu}}
    \end{flushleft}   

\newpage
\tableofcontents

\onecolumn
\setcounter{figure}{0}
\setcounter{section}{0}
\setcounter{table}{0}
\renewcommand{\thefigure}{S\arabic{figure}}
\renewcommand{\theequation}{S\arabic{equation}}

\section{NMIs referenced for ampere calibration and measurement capabilities}
The blue shaded region in Fig.~3b of the main text shows the typical calibration and measurement capabilities (CMCs) for electrical current offered by the following National Metrology Institutes (NMIs): Istituto Nazionale di Ricerca Metrologica (INRIM), Korea Research Institute of Standards and Science (KRISS), Van Swinden Laboratory (VSL), the Swiss Federal Institute of Metrology (METAS), the  National Institute of Metrology, China (NIMC), VTT Technical Research Centre of Finland's Centre for Metrology (VTT), Laboratoire national de métrologie et d'essais (LNE), Physikalisch-Technische Bundesanstalt (PTB), and Research Institutes of Sweden (RISE). The CMCs can be downloaded from the BIPM's key comparison database at \hyperlink{https://www.bipm.org/kcdb/}{https://www.bipm.org/kcdb/}.

\section{Indirect current realization schematic}
\begin{figure}[ht]
   \begin{centering}  \includegraphics[scale=0.7,keepaspectratio]{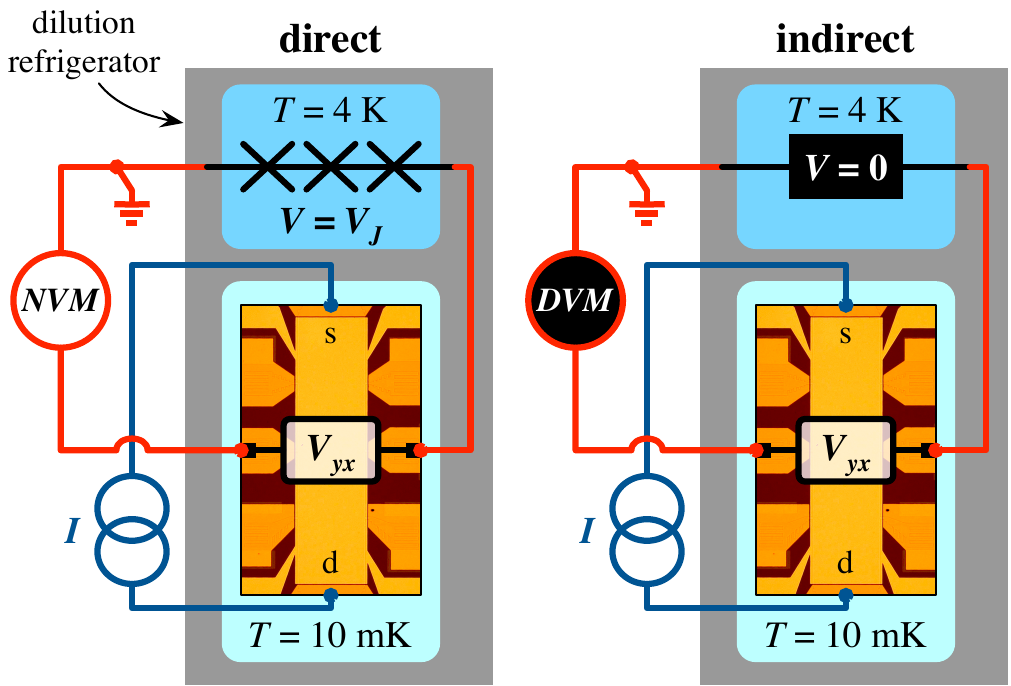}
       \caption{Simplified schematics depicting the difference between the direct (left) and indirect (right) quantum current measurement schemes. In direct mode, the programmable Josephson voltage standard (PJVS) is biased to output the voltage $V_J$ and the Hall voltage $V_{yx}$ is measured with a commercial nanovoltmeter. In indirect mode, the PJVS is disabled (i.e., the Josephson junction array becomes a superconducting short) and $V_{yx}$ is measured with a commercial, 8.5-digit digital voltmeter (DVM). For indirect mode, the current $I$ is determined using Ohm's law: $I_\mathrm{indirect} = V_{DVM}/R_K$.}
       \label{fig:SI:Iindir}
   \end{centering}
\end{figure}

Fig.~\ref{fig:SI:Iindir} compares the direct and indirect realizations of the ampere using the integrated prototype. The current $I$ was generated using the cryogenic current comparator's (CCC) digital current source and biases the quantum anomalous Hall resistor (QAHR) from source (s) to drain (d). See figure caption for additional details. 

\section{Interposer PCB}\label{sec:interposer}
\begin{figure}[H]
   \begin{centering}  \includegraphics[scale=0.6,keepaspectratio]{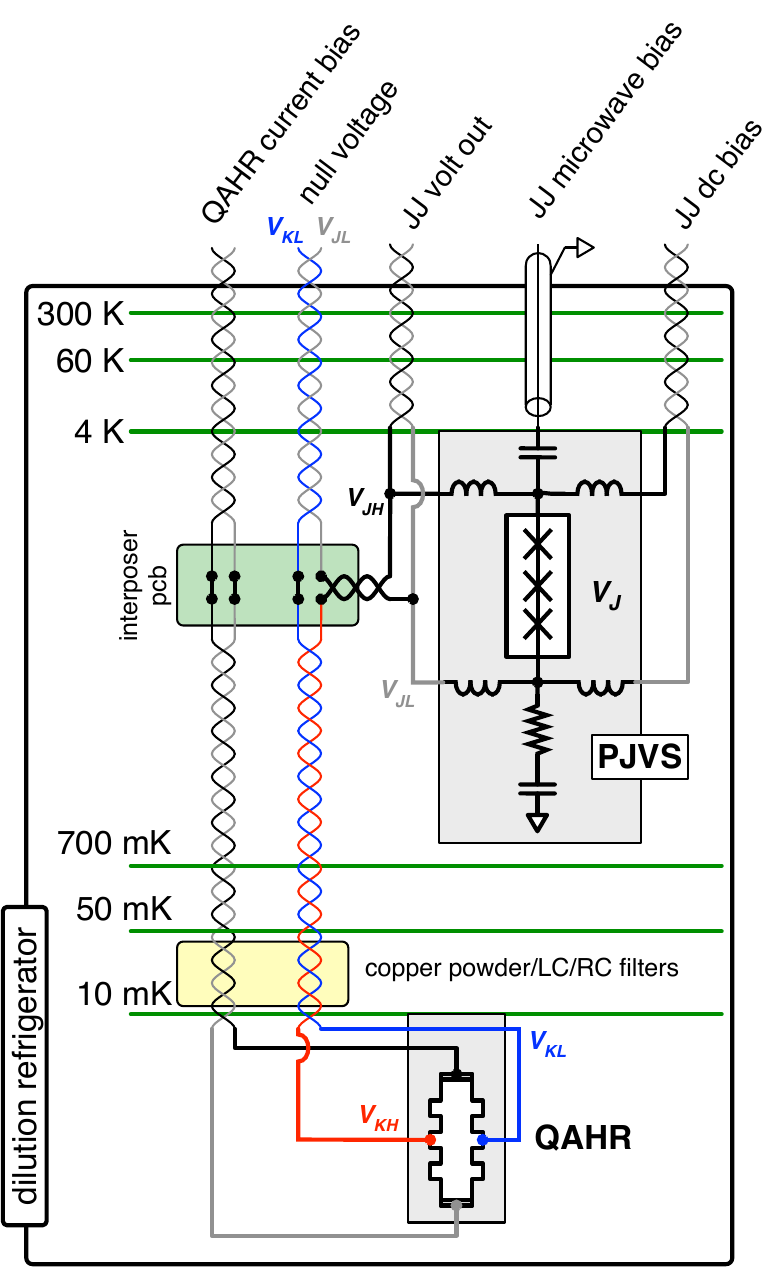}
       \caption{Wiring scheme for the unified instrument showing the locations of the PJVS, QAHR, filters, and interposer PCB (not to scale). The interposer facilitated interconnection of the PJVS and QAHR through the dilution refrigerator's connectorized, copper twisted pair cabling. The PJVS (mounted to 4~K stage) had on-chip inner-outer dc blocks, shown as capacitors in the schematic. The resistive element is a termination that is matched to the Josephson junction array's impedance (about 22~$\Omega$ at the end of the array). The inductive elements are superconducting low-pass filters that attenuate conducted leakage of the microwave bias to the precision dc leads. To further mitigate conducted noise, a surface mount, 10~nF capacitor (not shown) was soldered to the PCB between nodes $V_{JH}$ and $V_{JL}$. The QAHR was mounted on the end of a cold finger, which was clamped to the mixing chamber stage ($T =10$~mK). The twisted pair cabling to the QAHR was routed through copper powder and LC/RC filters at the mixing chamber stage. }
       \label{fig:SI:schematic_detail}
   \end{centering}
\end{figure}

To facilitate in situ characterization of the QAHR and to directly realize the ampere, a custom interposer PCB (printed circuit board) was inserted into the QAHR's twisted pair instrumentation cabling at the connectorized thermalization block mounted to the cryostat's 4~K stage. As depicted in Fig.~\ref{fig:SI:schematic_detail}, wire pairs for $I$ (and though not shown, $V_{xx}$) were passed straight through the interposer with surface mount zero-ohm jumpers between the input and output micro-D connector. In contrast, for $V_{yx}$, one of the wire pairs was routed off the interposer, through the voltage output taps of the PJVS, and then back to the output of the interposer. In this way, $V_{yx}$ and $V_J$ were always in series. During characterization of the QAHR, we disabled the PJVS (i.e., it is rendered a superconducting short on-chip) and thus the only added wire resistance was the short (20~cm) copper wire between the PJVS cryopackage and the interposer. 

To mitigate potential noise from the PJVS output when $V_J \neq 0$, we also soldered a 10~nF ceramic capacitor (NP0 dielectric) across the wire pair that connected the interposer and the PJVS. To check for possible failure, the capacitor was tested (after all measurements in the dilution refrigerator) by measuring its capacitance and leakage resistance at room temperature and while immersed in liquid helium. At both temperature extremes, leakage resistance was $>1$~G$\Omega$. Capacitance at 10~kHz changed by only 5~pF between the two temperatures.

\section{Uncertainty budget for the ampere}\label{sec:uncertainty}
\subsection{Uncertainty budget for \texorpdfstring{$I_\mathrm{direct}$}{Idirect} } 
\subsubsection*{Type A contribution}
\begin{itemize}
   \item  Null detector reading, $V_{\mathrm{null}}$: calculated as the standard deviation of means of multiple polarity reversals within a single experimental run. Lower values of current are naturally subject to larger amounts of relative dispersion given the lower signal to noise ratio and thus this contribution is largest for the lowest values of current. In cases where we have multiple experimental runs for the same current setpoint, we compute $I_\mathrm{direct}$ for each, and then take the weighted mean of the set to determine the values reported in the main text. The weighting we apply is simply the ratio of samples in a given run to the total samples in the set (at a given current). 
 \end{itemize}

\subsubsection*{Type B contributions}
\begin{itemize}
   \item Nanovoltmeter input impedance: When a digital voltmeter is used to measure $V_{yx}$, or the null between $V_{yx}$ and $V_J$, a voltage divider is formed between source impedances---such as the QAHR itself and resistive filters in the wiring---and the input impedance of the DVM. The same is true of any leakage resistance in the wiring. In a null measurement, the uncertainty contribution of the divider to the voltage estimate is diminished by the ratio of the null voltage to $V_J$ (or $V_{yx}$). The Keithley 2182A nanovoltmeter used for direct measurements of current was specified by its manufacturer to have an input impedance greater than 10~G$\Omega$. Assuming a worst-case null ratio of $10^{-3}$, the contribution of the  input impedance to the overall uncertainty would be less than 0.005~$\mu$A/A.  
   \item Nanovoltmeter gain and linearity: Similar to above, for direct/null measurements, the contribution of the nanovoltmeter's gain and nonlinearity to the overall uncertainty is diminished by the ratio of the null voltage to $V_J$. Even so, the nanovoltmeter used in these measurements was calibrated using NIST's conventional digital multimeter calibration service. For the 10~mV range, the gain error was 24~$\mu$V/V and the nonlinearity was less than 4~$\mu$V/V (note that the lowest  calibrated voltage possible was 1~mV). Assuming the same worst-case null ratio of $10^{-3}$, the contribution of the uncorrected gain error and nonlinearity to the uncertainty would be less than 0.03~$\mu$A/A. 
   \item PJVS microwave bias error: The laboratory where the DR was installed did not have access to a GPS-disciplined frequency standard. Instead, we used a commercial waveform generator (Keysight 33622A) with an onboard oven-controlled quartz crystal oscillator (OCXO) as our 10~MHz frequency reference. The microwave bias for the PJVS was phase locked to this standard for all measurements. We checked the frequency of the OCXO reference against a GPS-disciplined standard in a separate laboratory prior to and after our measurement campaign. Its frequency offset at both times was $-10 \pm 5 $~nHz/Hz. We also checked for phase locked loop deficiencies within the RF generator by measurements against a high frequency counter. No issues were found. We estimate this contribution to the overall uncertainty budget to be 0.01~$\mu$A/A. 
   \item $\delta R_{yx}$ combined uncertainty: $\delta R_{yx}$ is measured using a CCC as a ratio measurement against a standard resistor. The following are the sources of uncertainty for this measurement. 
   \begin{itemize} 
     \item Type A uncertainty related to the dispersion of the bridge voltage difference (BVD) measurements and its repeatability.
     \item Type B uncertainty due to the effects of PJVS microwave leakage on QAHR quantization. This contribution reflects our limited knowledge of the mechanisms by which Hall resistance quantization is impaired under microwave exposure. Since it is not possible to quantify and correct for this effect, we considered two approaches to evaluating its contribution to the overall uncertainty. One is to assume that the indirect and direct methods are equally valid realizations of the ampere and to perform a consensus analysis of the results. If the discrepancy is statistically significant, the consensus analysis will return a dark uncertainty, which represents an unresolved systematic. However, the consensus technique may overestimate the dark uncertainty if systematics are larger than the Type A (statistical) uncertainty. Given our knowledge that the microwave leakage effects are only present for direct realizations, the other approach is to favor the value achieved in the indirect realization and to assess the discrepancy as a bias. This is especially true for higher currents in which the discrepancy was much greater than the Type A uncertainty component. For our consensus analysis, we used NIST's online Consensus Builder (NICOB, \href{https://consensus.nist.gov}{consensus.nist.gov}). For the other approach, we defined the uncertainty as the root-mean-squared-error (RMSE) of each direct realization from the (favored) indirect realization. To avoid overestimating the uncertainty, we then took the minimum value among the two approaches (dark uncertainty or RMSE) for each current. Since the  uncertainties resulting from this analysis were within 1~$\mu$A/A of the magnitude of the corresponding discrepancies, for simplicity we assume that the contribution of the microwave leakage is equal to the magnitudes of the discrepancies. 
     \item Type B uncertainty related to the SQUIDs rectification of noise causing an offset in the BVD measurement. This can be measured by measuring the BVD with the current sources off.
     \item Compensation network drift: Type B uncertainty due to the drift of the compensation network. The compensation network is a series-parallel combination of resistors which act to bring the BVD as close to zero as possible. Typically this network is calibrated once a year. 
     \item Standard resistor: Combined uncertainty of the standard resistor used to characterize the QAH device. We used a calibrated 100~$\Omega$ standard resistor as the reference for these measurements with an estimated uncertainty of 3~n$\Omega/ \Omega$\ (\textit{k} = 1)
     \item Leakage: From room temperature leakage measurements, we estimated electric isolation of the dc wiring and connections to be on the order of 100~\si{\giga\ohm} (resistance to ground and intra-pair resistance). This results in a relative error of a few parts in $10^{7}$ and mostly limits the precision of the resistance measurements. 
     \item CCC winding ratio error: The windings of the CCC used in this work were set by connecting a series of windings with different turns in an arithmetic combination. For ratios other than 1:1, the errors must be determined by a step-up calibration procedure based on $1:1$ intercomparisons of all windings~\cite{grohmann1979,lu2020} This error is small and on the order of a few parts in $10^{-10}$ but must be taken into account.
   \end{itemize}
 \end{itemize}

\subsection{Uncertainty budget for \texorpdfstring{$I_\mathrm{indirect}$}{Iindirect}}
The uncertainty budget for $I_{\mathrm{indirect}}$ includes a Type A contribution for the Keysight 3458A digital voltmeter (DVM) readings, which ranged from 0.60~$\mu$A$/$A at 252~nA to 31.0~$\mu$A$/$A at 9.33~nA. As discussed in the main text, we did not correct for the DVM's raw error (i.e., the discrepancy between its reading and a known applied voltage), as it was always less than the expanded ($k=2$) Type A uncertainty. Instead we assigned a relative Type B uncertainty contribution for the DVM of 1~$\mu$A$/$A. We also estimated a negligible relative Type B uncertainty of 0.1~$\mu$A/A due to the measured input impedance of the DVM ($Z_\mathrm{in}$ of order 1~T$\Omega$ on 100~mV range). Finally, we used the same overall uncertainty (1~$\mu$A$/$A) for the QAHR from the direct mode uncertainty budget. 

\section{Effect of magnetic field on PJVS}
To assess the effect of magnetic field on PJVS' performance, we swept the field in the bore of the solenoid up to 1~T and measured the stray field outside the vacuum envelope, but near the PJVS (the radial separation between PJVS cryopackage and probe was approximately 10~cm). For a 1~T solenoid field, we measured a stray field magnitude of approximately 1~mT near the PJVS, a value which is consistent with the field map of the bare solenoid (with integrated bucking coil) and the passive shielding added by the fridge's manufacturer. No additional magnetic shielding was installed around the PJVS cryopackage. During the sweep, we also monitored the critical current and Shapiro step width of one of the 8 400-junction subarrays for a bias frequency of 9.7 GHz. We found no observable change in the PJVS critical current or Shapiro step width at the maximum field. 

To allow comparison to historical performance at liquid helium temperatures, we heated the fridge's 4~K stage to approximately 4.2~K. The critical current and Shapiro step widths were comparable to those measured in liquid helium: $I_c = 6.4$~mA, $\Delta I_{SS} = 2$~mA for the fridge at 9.7~GHz and approximately 4~K, compared with $I_c = 6.2$~mA, $\Delta I_{SS} = 2.5$~mA at 18~GHz in liquid helium. The difference in step width could be due to the response of the junction array vs frequency or the fact that the power amplifier used in this work has a lower maximum output power. During the field sweep, we did observe a slight decrease in critical current each time the field was stepped. Since we also observed increases in the temperature of the 4~K stage at each field increment, we attribute the reduction in critical current to changes in temperature, rather than a direct result of the field itself. 

\section{Lock-in amplifier qualification of candidate Cr-BST films}
\label{sec:lock-in msmts}
Prior to metrological characterization, candidate Cr-BST films were qualified via low-frequency ac transport measurements. For these measurements, 100~$\mu$m wide Hall bar heterostructures were fabricated from the same 6-QL Cr-BST thin-film described in the main text. We used a current with an rms amplitude of 5~nA to bias the Hall bar. We measured the longitudinal voltage $V_{xx}$ and Hall (transverse) voltage $V_{yx}$ with a lock-in amplifier as a function of top gate voltage $V_g$ and temperature $T$. From these voltages, the longitudinal and transverse resistivities can be computed according to, 
\begin{equation}\label{eq:rho_xx_rho_yx}
 \rho_{xx} = \frac{w}{l}\frac{V_{xx}}{I}, \; \rho_{yx} = \frac{V_{yx}}{I}, 
\end{equation}
where $w$ is the transverse width of the Hall bar device, $l$ is the center-to-center distance between the voltage probes used to measure $V_{xx}$, and $I$ is the longitudinal bias current. The conductivity tensor components $\sigma_{xx}$ and $\sigma_{xy}$ can then be derived from the resistivities as, 
\begin{equation}
    \sigma_{xx} = \frac{\rho_{xx}}{\rho_{xx}^2 + \rho_{yx}^2}, \; \sigma_{xy} = \frac{\rho_{yx}}{\rho_{xx}^2 + \rho_{yx}^2}. 
\end{equation}
Measurements of $\rho_{xx}$ as a function of gate voltage and temperature are often used to adjust the Fermi level to the center of the magnetic exchange gap, where the QAH state is most robust. Empirically, we observed that the native Fermi level in our Cr-BST films shifts over time, with optimal gate voltages moving towards more positive values: from $V_g = -1.35$~V to near $V_g = 0$~V over a period of about two years, which is approximately the span between the lock-in qualification at Stanford University and the metrological characterization at NIST. 

\begin{figure}[H]
    \begin{centering} 
    \includegraphics[scale=0.75]{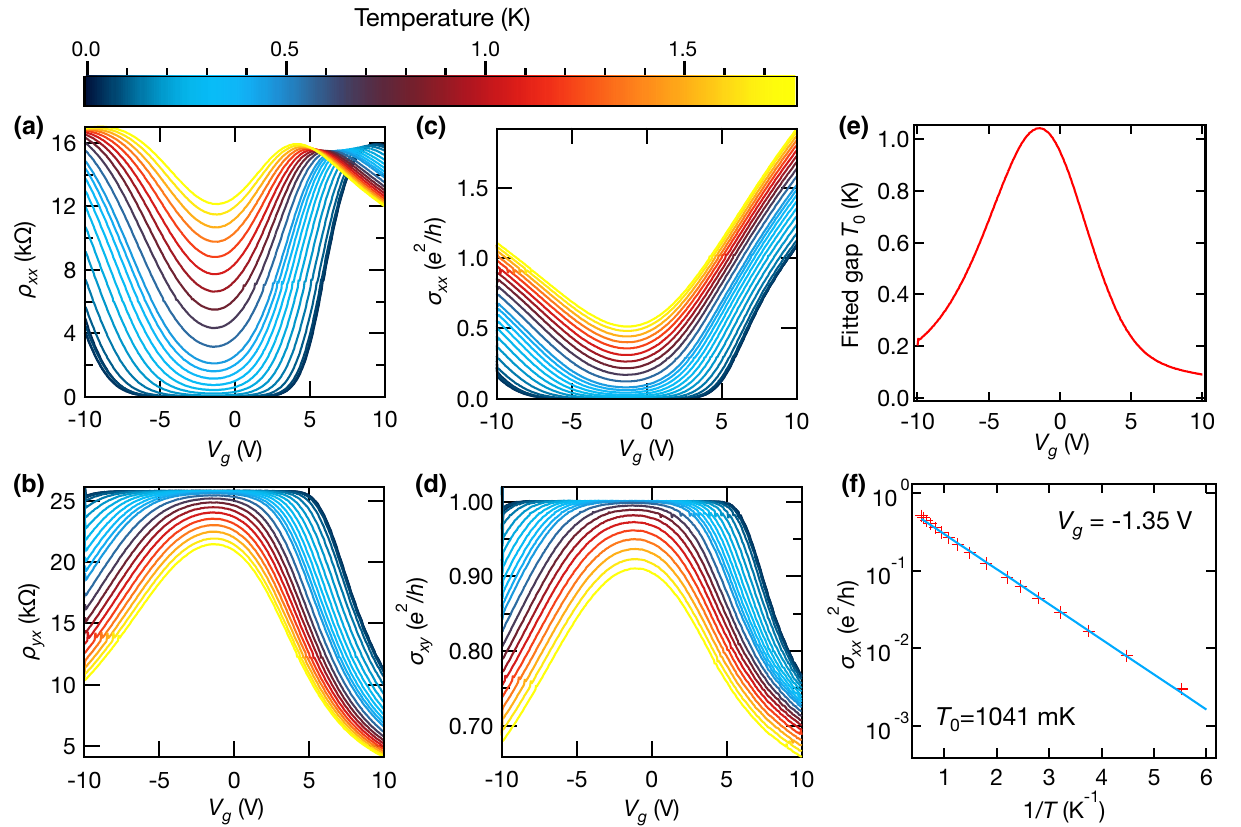}
        \caption{Lock-in characterization of a Hall bar device fabricated from the same Cr-BST film used to make the QAH device used in the main text. The longitudinal resistivity \textbf{(a)} and longitudinal conductivity \textbf{(c)}, as well as Hall resistivity \textbf{(b)} and Hall conductivity \textbf{(d)} are plotted as a function of top gate bias voltage $V_g$ for different values of the mixing chamber temperature $T$. Base temperature was approximately $30$~mK. \textbf{(e)} Fitted thermal activation temperature scale $T_0$ as a function of $V_g$ shows that by modulating the potential of the top gate we are able to tune the Fermi level through the center of the magnetic exchange gap. \textbf{(f)} The longitudinal conductivity $\sigma_{xx}$ measured at $V_g = -1.35$~V is plotted as a function of inverse temperature on a $\log$-$\log$ scale. A fit to thermally activated conduction $\sigma_{xx} \propto e^{-T_0/T}$ for $T > 160$~mK is shown as a solid line. Empirical evidence suggests that the film used in the main text should have a magnetic exchange gap comparable to magnitude that shown in \textbf{(e)} but shifted such that $T_0^{\mathrm{max}}$ is near $V_g = 0$~V.}
        \label{fig.lt3745}
    \end{centering}
\end{figure}

Lock-in measurement results are shown in Fig.~\ref{fig.lt3745}a--d. At base temperature we observe a large plateau spanning $V_g \approx -6$~V to $V_g \approx 3 $~V where $\rho_{xx} = \sigma_{xx} \approx 0$, $\rho_{yx} = h/e^2$, and $\sigma_{xy} =e^2/h$ indicating that the Fermi level is positioned within the magnetic exchange gap. Beyond these gate voltages, increased dissipation and deviations from quantization become apparent as the Fermi level approaches the dissipative surface states. Within the plateau, dissipation appears thermally activated with $\sigma_{xx} \propto e^{-T_0/T}$ above $T \approx 160$~mK as seen in Fig.~\ref{fig.lt3745}d. The fitted thermal activation temperature scale $T_0$ peaks at $V_g = -1.35$~V with $T_0 = 1041$~mK as shown in Fig.~\ref{fig.lt3745}e.

\section{The CCC's digital current source}\label{sec. IGUI and I}
We used the cryogenic current comparator's (CCC) digital current source to bias the QAHR during validation of its resistance and for realizing a measurement of the ampere. The current source is based on a 16-bit digital-to-analog converter (DAC)\cite{gotz2009}. As a result, $I$ could not be changed continuously but had to be incremented in pre-defined steps using a graphical user interface (GUI). In the main paper, the nominal currents $I_\mathrm{nom}$ are referenced to 3 significant figures as an aid to the reader. Additionally, any currents that were within 500~$\mu$A$/$A of one another were grouped (using the arithmetic mean) into a single data point in plots for clarity. Table~\ref{tab.All_I_comps} below shows the corresponding values of current used in this work. 

We typically found that the measured current---using either direct or indirect mode---was slightly lower than that indicated by the CCC's GUI. The largest disagreement between $I_\mathrm{GUI}$ and $I_\mathrm{direct}$ or $I_\mathrm{indirect}$ was roughly -300~$\mu$A$/$A for $I = 83.9$~nA. As described in the main text, when operating the CCC as a resistance bridge (it's primary function), only the ratio of the currents in the two arms is critical and is kept fixed through feedback from the SQUID. Therefore, the deviations of the current source's measured output from the GUI's values are not indicative of any defect with the CCC.

\begin{table}[h]
\centering
\caption{\label{tab.All_I_comps} Comparison of $I_\mathrm{nom}$, $I_\mathrm{GUI}$, $I_{\mathrm{direct}}$, and $I_{\mathrm{indirect}}$. First column shows the nominal values used in the main text. Second column shows the respective GUI values,   $I_\mathrm{GUI}$. Third and fourth column: measured values of $I_\mathrm{direct}$ and $I_\mathrm{indirect}$ respectively.}
\begin{tabular}{|| c | c | c | c ||} 
\hline
$I_\mathrm{nom}$ (nA) & $I_{\mathrm{GUI}}$ (nA)& $I_{\mathrm{direct}_\textrm{ }}^{\mathrm{avg}}$ (nA)&  $I_{\mathrm{indirect}}^{\mathrm{avg}^{\textrm{  }^{\textrm{ }}  }}$ (nA)\\
\hline
9.33& 9.329500 & 9.328131 & 9.328337 \\
\hline
28.0 & 27.97940  & 27.97921  & 27.97988 \\
\hline
28.0 & 27.98850  &  27.98659 & 27.98659 \\
\hline
53.8 & 53.78890 & 53.78686 & 53.78732 \\
\hline
83.9 & 83.92895 & 83.91331 & 83.91356 \\
\hline
83.9 & 83.93810  & 83.92265 & 83.92263 \\
\hline
83.9 & 83.94730 & 83.92946 & 83.92964 \\
\hline
132 & 132.3893 & 132.3733 & 132.3750 \\
\hline
132& 132.3984  & 132.3832 & 132.3843 \\
\hline
252& 251.8235 &  251.7836 & 251.7877 \\
\hline
252& 251.8326 & 251.7938 & 251.7972 \\
\hline
\end{tabular}
\end{table}

\section{Corresponding values of \texorpdfstring{$I$, $N$, $f$ and $V_J$}{parameters}}\label{sec:corresponding n,f,i}
As discussed in the main text, the PJVS microwave bias had a measurable impact on the longitudinal resistance of the QAHR. We selected three nominal PJVS bias frequencies, which we found to have limited impact on quantization of the QAHR (see Section~\ref{sec: microwave heating}). The nominal frequencies were $f_\mathrm{0} = 9.701$~GHz, $f_1 = 18.65$~GHz, and $f_2 = 15.30$~GHz. The PJVS output voltage depends on the number of junctions $N$ and the bias frequency $f$ according to, 
\begin{equation}\label{eq. VJ}
    V_J = \frac{Nf}{K_J}, 
\end{equation} 
where $K_J = 2e/h$ is the Josephson constant. As discussed in the Methods section of the main text, individual JJs are grouped into subarrays on the PJVS and so $N$ can take corresponding values of 12, 36, 108, and 324. To realize the ampere in direct (null) mode, ideally one would adjust $N$ and $f$ to achieve a cancellation of the Hall voltage $V_{yx} = I R_K$ for any applied current $I$. In our prototype, the constraints on $f$ and $N$ limited the possible values of $V_J$, and in turn the values of the ampere that we could realize directly. The corresponding values of the various parameters are depicted in Table~\ref{tab.PJVS to I}.

\begin{table}[h]
\centering
\caption{\label{tab.PJVS to I} PJVS settings for each QAH bias current $I$ (first column). Second column: $f$. Third column: $N$. Fourth column: resulting PJVS output voltage calculated using Equation~\ref{eq. VJ}.}
\begin{tabular}{||c | c | c | c  ||} 
\hline
$I$ (nA) & $f$ & N & $V_J$~ (mV) \\
\hline
9.33 & $f_0$ & 12 & 0.2407232  \\
\hline
28.0 & $f_0$&  36  & 0.7221697\\
\hline
53.8 & $f_1$ & 36 & 1.388444\\
 \hline
83.9 &  $f_0$ & 108 & 2.166442\\
\hline
132 & $f_2$& 108 & 3.416889 \\
\hline
252 &  $f_0$ & 324  & 6.499527\\
\hline
\end{tabular}
\end{table}

\section{Parameter response testing}\label{sec:response_testing}
During preliminary indirect experiments, it appeared that, for certain currents, the measured value would occasionally exhibit discrete step changes in subsequent runs. For example, if we measured current at 10 distinct GUI values and then repeated measurements of those same values at a later time, a subset would shift. Since the behavior only affected some, and not all, currents, it is distinct from an overall gain drift of the instrumentation. To better understand the impact of shifts in one or more of the selected currents, we evaluated the null voltage's response to \emph{changes} in the bias parameters $I$ and $f$. The wiring of the two quantum standards in our unified instrument followed a convention such that the null voltage is given by, 
\begin{equation}
    \label{null_voltage}
    V_\mathrm{null} = V_J - I_\mathrm{direct} R_{yx}.  
\end{equation}
Thus, if $I$ and $f$ (and therefore $V_J$) were adjusted by known amounts, we expect the null voltage to respond in a perfectly linear manner to those changes. Specifically, we wish to check the validity of the following relationship, which follows from Equation~\ref{null_voltage}, 
\begin{equation}
    \label{functional_test}
    \frac{\delta V_\mathrm{null}}{V_{J,1}} = \frac{\delta f}{f_1} - \left(\frac{\delta I}{I_1}+\frac{\delta R_{yx}}{R_{yx,1}}\right),  
\end{equation}
where the subscript `1' in the denominators indicates the parameter value prior to any shift and we assume $V_\mathrm{null}/(IR_{yx}) \ll 1$. For the moment, we further assume that the relative change in the QAHR's Hall resistance, $\delta R_{yx}/R_{yx}$, is negligible between measurements under the two sets of bias parameters. We then evaluate Equation~\ref{functional_test} using the same raw data that Figure~3 in the main text is based upon. The validation of Equation~\ref{functional_test} is depicted in Table~\ref{functional_test_table}.  
\begin{table}[ht]
\centering
\caption{Validation of Equation~\ref{functional_test} in direct mode. For a given nominal current $I_\mathrm{nom}$, PJVS bias frequency $f$ and/or QAHR bias current $I$ were adjusted by the relative amount indicated and the relative change in null voltage $V_\mathrm{null}$ was computed from measurements. The expected and measured changes in $V_\mathrm{null}$ and their difference $\Delta$ are shown (all scaled by the initial Josephson voltage $V_J$). $\Delta$ may be compared to the type A uncertainty ($k=1$) for a single direct mode measurement, shown in the right column.}
\label{functional_test_table}
\begin{tabular}{cccccccc}
\toprule
\multirow{2}{*}{$I_\mathrm{nom}$ (nA)} & \multirow{2}{*}{Parameter} & \multirow{2}{*}{$\frac{\delta f}{f}$ $\left(\frac{\mu \textrm{Hz}}{\textrm{Hz}}\right)$} & \multirow{2}{*}{$\frac{\delta I}{I}$ $\left(\frac{\mu \textrm{A}}{\textrm{A}}\right)$} & \multicolumn{3}{c}{$\frac{\delta V_\mathrm{null}}{V_J}$ $\left(\frac{\mu \textrm{V}}{\textrm{V}}\right)$} & \multirow{2}{*}{$U_\mathrm{null,A}$ $\left(\frac{\mu \textrm{V}}{\textrm{V}}\right)$} \\ 
 &  &  &  & \multicolumn{1}{c}{exp.} & \multicolumn{1}{c}{meas.} & \multicolumn{1}{c}{$\Delta$} &  \\
\midrule
28.0 & $I$ & — & 240 & -240 & -238 & 1.7 & 12 \\
83.9 & $I$ & — & 83.5 & -83.5 & -81.1 & 2.4 & 3.8 \\
83.9 & $f$, $I$ & 31.0 & 108 & -77.2 & -80.3 & -3.2 & 3.8 \\
83.9 & $f$, $I$ & 31.0 & 192 & -161 & -162 & -0.8 & 3.8 \\
132 & $I$ & — & 70.3 & -70.3 & -72.6 & -2.3 & 2.2 \\
132 & $f$ & 132 & — & 132 & 135 & 3.1 & 2.2 \\ 
\bottomrule 
\end{tabular}
\end{table}

In evaluating Equation~\ref{functional_test}, the experimental changes to $I$ and $f$ were all made while operating in direct mode. However, recall that for $I$, we can only change the CCC's GUI setting, which does not perfectly represent the actual current sourced. Therefore, the \emph{values} for $I$ corresponding to those settings were determined from the indirect mode's quasi-calibration. If one of the two currents in each response test had in fact shifted, the data should not satisfy Equation~\ref{functional_test}. Instead, the deviation $\Delta$ between the expected and measured changes in $V_\mathrm{null}$ are comparable to the type A uncertainty (and all fall within expanded uncertainty). Thus, we can conclude either (a) there was no significant change in the current sourced between direct and indirect measurements or (b) the pairs of currents changed by the same amount. Based on analysis of our indirect measurements over time (see, e.g., Fig.~4a in the main text) and an evaluation of the possible errors in the current source's DAC (see section \ref{sec:uncertainty}), we find no compelling evidence to support the latter scenario.  

As discussed in the main text, it is likely that the QAHR's Hall resistance $R_{yx}$ varied during direct measurements as a result of the microwave leakage from the PJVS. Thus, an uncontrolled variation in $R_{yx}$ may be confounded with the intentional change in $I$. And yet, it seems unlikely that the current and Hall resistance would both vary in such a way as to yield $\Delta \approx 0$ for five different comparisons. Indirect measurements before and after the direct campaign strongly suggest that the QAHR did not suffer a \emph{permanent} loss of quantization—at least not one large enough to explain the disagreements observed (again, see Fig.~4a of the main text). However, we cannot rule out ephemeral quantization changes due to the cycling of the microwave bias and varying frequencies during direct realizations (see also section~\ref{sec: microwave heating}).

\section{Realization of the ampere with a second QAHR}
After the initial measurement campaign described in the main text, we repeated some of our SI unit realizations with a second QAHR. The second device was fabricated on a different Cr-BST film and had a smaller geometry (150~$\mu$m vs 300~$\mu$m width) and a lower thermal activation temperature. Unsurprisingly, the second device did not perform as well in terms of its susceptibility to microwave leakage. 

\begin{figure}[H]
   \centering
  \includegraphics[scale=0.8]{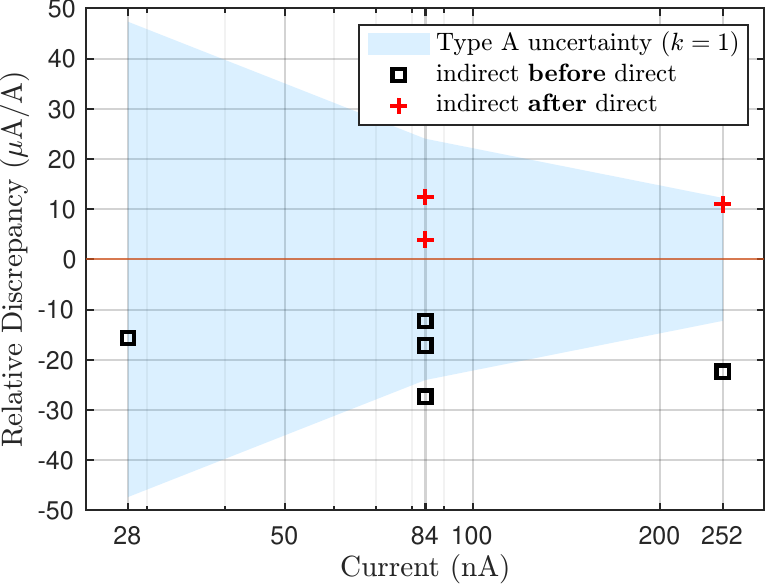}
\caption{Disagreement between direct and indirect realizations of the ampere with a second QAHR. Measurements were performed over approximately five days. The Type A uncertainty of the measurements is shown as the light blue shaded region. Black squares depict the disagreement between the direct realizations and only the indirect realizations that occurred just prior in time. Red crosses show disagreements when comparing the same direct data to the indirect realizations later in time. The upward shift suggests a permanent loss of quantization in the QAHR as a result of microwave leakage from the PJVS.} 
   \label{fig.mew}
\end{figure}

The second device was measured with the same experimental arrangement described in the main text, with the exception of different PJVS bias electronics that had greater isolation to ground. Fig.~\ref{fig.mew} shows the disagreement between direct and indirect realizations of the ampere using the second QAHR. Here, the different symbols (black squares, red crosses) are used to indicate two sets of comparisons. Again, due to limited time allotted on the cryostat, first we performed indirect realizations for the currents shown, then we performed direct realizations of the same currents, and finally we repeated the indirect measurements for as many points as possible. 

The black squares in Fig.~\ref{fig.mew} show the disagreement between the first set of indirect realizations and the direct realizations. Multiple symbols shown near the same current indicate realizations at different points near the nominal value. The disagreement among all currents was in the range (-30 to -10)~$\mu$A/A, which is comparable to that with the first QAHR. Curiously, if we instead compare the direct realizations with the second set of indirect data (red crosses), an upward shift of (20 to 30)~$\mu$A/A is evident. We attribute this shift to a permanent loss of quantization in the QAHR as a result of microwave leakage from the PJVS (see section \ref{sec: microwave heating} for additional details). However, since the Type A uncertainty on this device was comparable to the magnitude of the shift, we caution that this attribution is provisional. 

\newpage
\section{Impact of PJVS microwave leakage on QAHR}
\label{sec: microwave heating}
The longitudinal and Hall resistivities of the QAHR are strongly dependent on temperature, as seen in Section~\ref{sec:lock-in msmts} where $\rho_{xx}$ increased by two orders of magnitude between $T=30$~mK and $T=1.7$~K while $\rho_{yx}$ decreased by roughly 20\% (at $V_g = -1.35$~V). Anticipating the QAHR's sensitivity to unwanted heating from thermal noise, we installed cryogenic low-pass filters (QDevil QFilter-II) in the cryostat. Likewise, a radiation shield that envelopes all thermal stages at or below 700~mK was expected to provide additional protection from stray environmental noise. Furthermore, the vastly different temperature regimes of the two quantum standards (10~mK and 4~K) enforced a spatial separation of about 1~m, due to the thermal stage arrangement in the cryostat. Despite these mitigations, we observed changes in the QAHR's quantization when the PJVS's microwave bias was active. 

While we do not yet know the precise mechanism that results in quantization changes (e.g., purely thermal and/or photon-assisted tunneling between charge puddles in the QAHR bulk), subsequent checks with the QAHR and PJVS disconnected from each other indicate that the unwanted coupling is indeed radiative. That is, galvanic interconnection of the standards through the twisted-pair cabling is not necessary to replicate  microwave-induced breakdown of the QAHR. We believe this finding will help inform the design of future prototypes by focusing on improved shielding and the reduction of cavity modes. 

\begin{figure}[ht]
    \begin{centering}  \includegraphics[width=\textwidth,keepaspectratio]{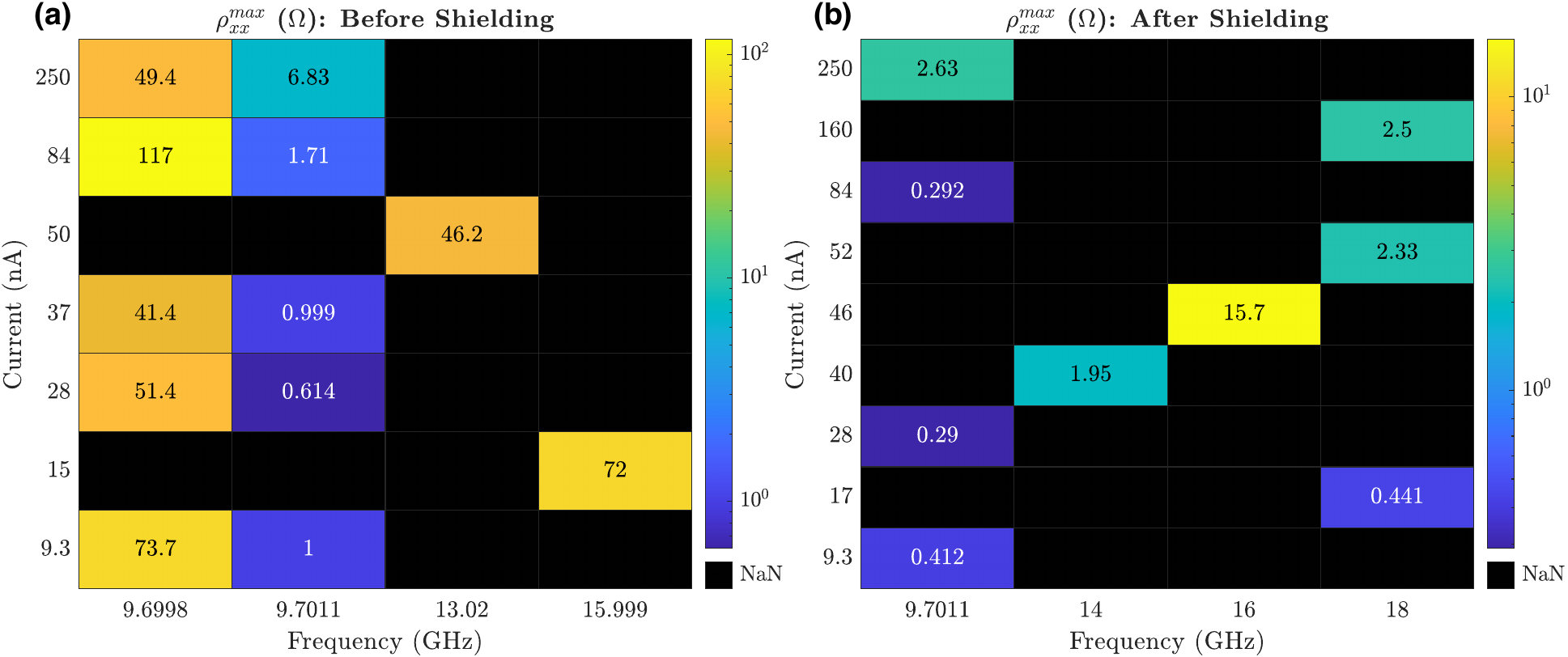}
        \caption{Heat maps of the maximum measured value of $\rho_{xx}$ in $\Omega$ (a) without and (b) with RF shielding installed around the PJVS. Note that $x$- and $y$-axis values, as well as logarithmic colormap range, are different between the two panels. Each $(I,f)$ grouping is based on at least 10 measurements. Both panels highlight our observation that a microwave bias at $f_0=9.7011$~GHz had the smallest impact on $\rho_{xx}$.}
        \label{fig.rho_xx_shielding_heatmap}
    \end{centering}
\end{figure} 

In preliminary experiments without radio frequency (RF) shielding around the PJVS, we noticed temperature excursions on the cold finger as high as 100~mK (from the 10~mK base temperature). We could not distinguish whether the stray radiation was heating the cold finger or heating the resistance thermometer mounted to it (e.g., by coupling to the thermometer wiring). However, we did rule out the possibility that the disagreement was the result of RF rectification at the inputs of the Keithley nanovoltmeter and the Keysight voltmeter. Fig.~\ref{fig.rho_xx_shielding_heatmap} illustrates the dissipation reductions we observed after adding RF shielding as a function of QAHR longitudinal current and PJVS microwave bias frequency. Fig.~\ref{fig.rho_xx_shielding_heatmap}a shows the maximum value of longitudinal resistivity $\rho_{xx}^{max}$ from measurements without RF shielding. Microwave-induced dissipation was lowest for the frequency $f_0 = 9.7011$~GHz, which motivated the choice to focus on that particular frequency for most realizations. Fig.~\ref{fig.rho_xx_shielding_heatmap}b shows the same quantity with the shield installed around the PJVS. At $f = f_0$, the shielding reduced dissipation by at least half for the currents tested. Additionally, the shielding made it possible to attempt realizations at other frequencies. However, for certain frequencies (e.g., $f=16$~GHz), dissipation was still too high, even with the RF shield. 

\begin{figure}[H]
    \begin{centering}  \includegraphics[width=\textwidth,keepaspectratio]{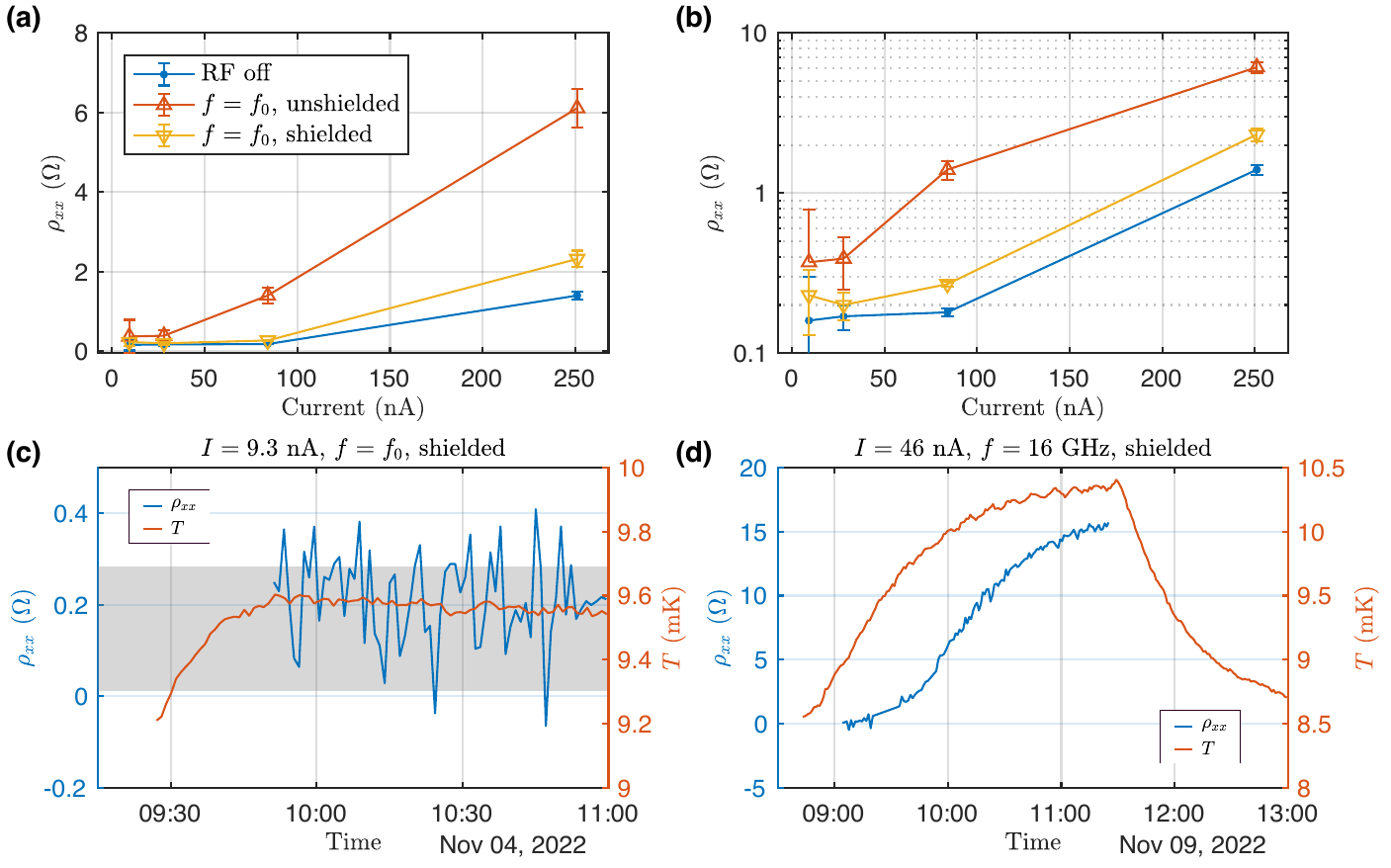}
        \caption{Impact of PJVS  microwave leakage on the electron temperature of the QAHR (via its longitudinal resistivity $\rho_{xx}$). \textbf{(a)} $\rho_{xx}$ as a function of bias current $I$ for different microwave exposures: No microwave bias (blue dots), microwave bias enabled and PJVS unshielded (orange upward triangles), and microwave bias enabled and PJVS shielded (yellow downward triangles). Each symbol is the average of at least 30 observations and the error bars represent the Type A uncertainty of those observations. \textbf{(b)} The same data as in \textbf{(a)} but shown on a semi-log scale (see legend in \textbf{(a)}). Error bars represent the standard deviation of the measurements. As in the main text, $f_0 = 9.7011$~GHz. \textbf{(c)} Time trace of $\rho_{xx}$ at $I=9.3$~nA and mixing chamber temperature $T$, shortly after enabling microwave bias at 9:27~AM local time. Here, $f=f_0$ and the PJVS was shielded. The grey rectangular patch shows the range of $\rho_{xx}$ for the same current, but with microwave bias disabled. \textbf{(d)} Another $(\rho_{xx},T)$ time trace for $I=46$~nA and $f=16$~GHz, showing that certain frequencies may produce significant heating, even when PJVS was shielded. The microwave bias was first turned on at 8:45~AM. Although we do not have data for 46~nA at $f_0$, we can estimate from \textbf{(b)} that the corresponding $\rho_{xx} \approx 0.25$~$\Omega$, 1/50th of the value at the end of the trace in \textbf{(d)}. For all panels, the microwave power was $P=0$~dBm.}
        \label{fig.rf heating}
    \end{centering}
\end{figure}

Figures~\ref{fig.rf heating}a,b show alternative illustrations of the microwave leakage effect at $f = f_0$ (see caption for details). As noted in the main text, the additional shielding helped, but did not completely eliminate microwave-induced dissipation in the QAHR. For example, Fig.~\ref{fig.rf heating}c suggests that, at least for $f = f_0$ and PJVS shielded, the effect of leakage on $\rho_{xx}$ is minimal. While it is evident that the leakage also heats the mixing chamber stage (as measured by its ruthenium oxide thermometer), we do not believe that the increases in $\rho_{xx}$ are simply the result of increased lattice temperature. Fig.~\ref{fig.rf heating}d shows, as noted above, that the wrong choice of bias frequency results in significant dissipation and that the timescale for such effects can be hours. 

The evolution of longitudinal and Hall resistivities after application of the PJVS microwave bias has implications for our direct realizations of the ampere. Our measurements of the ampere were computer automated to run unattended. As part of that protocol, the microwave bias was enabled prior to the start of measurements and disabled when measurements completed to protect instrumentation and devices in the event of facilities outages. The limited time available on the dilution refrigerator precluded waiting hours prior to each measurement to allow the QAHR to stabilize. As a result, it is likely that many of our ampere realizations occurred while the QAHR was still relaxing toward a new steady state (in the presence of microwave-induced dissipation). 

\begin{figure}[H]
   \centering
  \includegraphics[scale=0.4,keepaspectratio]{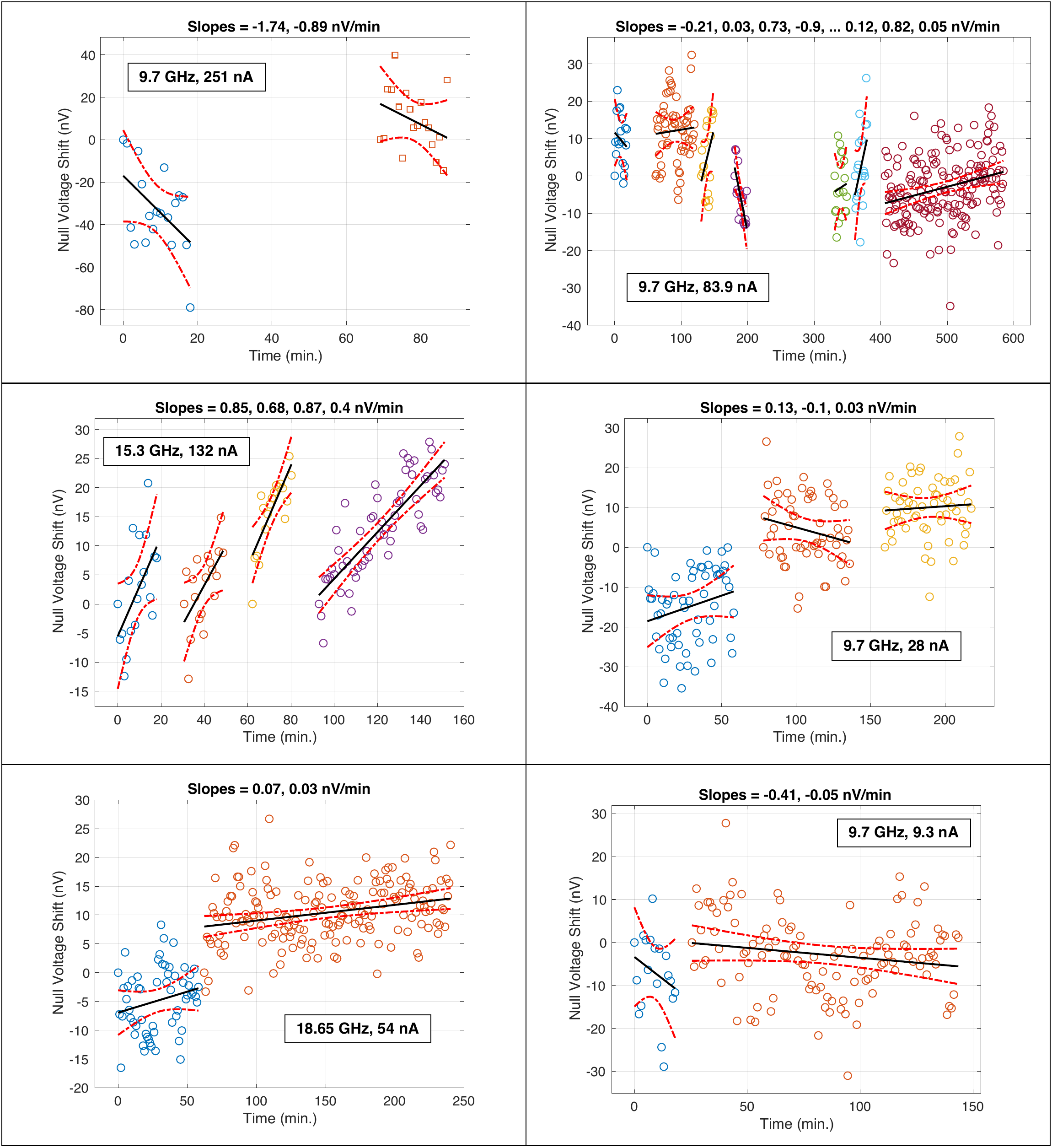}
\caption{Variation with time of null voltage $V_\mathrm{null}$ during direct realizations of the ampere. For each of the six panels, the clusters of data symbols show the change in $V_\mathrm{null}$ relative to the data symbol at the start of that run. Black solid lines show a linear fit to the data, while the dashed red lines show the 95\% CI for the fit. The different symbol colors are intended as a guide so the reader can distinguish individual runs; they have no relationship between the panels. The panels are ordered according to the sequence in time that the measurements were performed, first increasing from left to right, and then from top to bottom. Text boxes show the nominal frequency and current for each panel. Note that both frequency and the current may vary about the nominal value between runs. The titles of each plot show the numerical slope of the fits for each run. Additional discussion can be found in the text. } 
   \label{null_drift_analysis}
\end{figure}

Such a hypothesis is supported by analyzing drift in the raw data from our direct realizations of the ampere. The panels in Fig.~\ref{null_drift_analysis} show the evolution in time of the null voltage from the start of each run. The microwave bias was enabled just prior. For example, the top left panel (9.7~GHz, 251~nA) shows two runs, each lasting approximately 20~min and separated by 50~min. The individual data symbols show the shift in null voltage relative to the very first symbol of each run. The black solid line shows a linear fit to the data and the red dashed lines show the 95\% confidence intervals (CIs) for the fit. The slopes of the fits for each run are shown in the title for each panel. Under the assumption that the shift in null voltage is only the result of a change in $R_{yx}$, we can use ~Equation~\ref{null_voltage} to relate the two quantities: 
\begin{equation}
    \label{null_voltage_Ryx_shift}
    \frac{1}{R_{yx}}\frac{d R_{yx}}{dt} = -\frac{1}{V_J}\frac{d V_\mathrm{null}}{dt}.  
\end{equation}

As noted above, when the QAHR is exposed to elevated temperatures, $\rho_{xx}$ increases while $R_{yx}$ decreases. In our precision measurements of $R_{yx}$ we found evidence which suggests that, at least for some frequencies, $R_{yx}$ initially decreases immediately following microwave exposure, but may subsequently recover quantization (i.e., $R_{yx}$ increases) after some time. Taking again the first run of the top left panel of Fig.~\ref{null_drift_analysis} as an example ($V_J=6.499527$~mV), we use Equation~\ref{null_voltage_Ryx_shift} to compute a rate of change in $R_{yx}$ of $+0.27$~$\mu\Omega/\Omega/$min. Integrating over the 20~min duration of that run, we derive an overall change of $+5.4$~$\mu\Omega/\Omega$. While the magnitude of that change cannot completely account for the direct-indirect disagreement and the range of slopes within the 95\% CIs span both negative and positive values, it is consistent with a picture in which the QAHR suffers an initial loss and subsequent recovery of quantization. 

For the other panels in Fig.~\ref{null_drift_analysis}, such a picture is less obvious. However, we highlight two salient points. First, for $f$ near 9.7~GHz and 18.65~GHz, the slope of the fit generally approaches zero with each successive run, as long as they are within about an hour of each other. This behavior seems to suggest that quantization recovery is not lost immediately upon disabling the microwave bias. Second, for $f$ around 15.3~GHz, the slopes are consistently positive for all runs, which suggests that the QAHR diverges from quantization (i.e., $R_{yx}$ decreases) as the duration of microwave exposure increases and does not recover until long after exposure ends. This is consistent with the relatively high $\rho_{xx}$ measured for frequencies between (14--16)~GHz (see Fig.~\ref{fig.rf heating}). 

\section{Allan deviation measurements of the nanovoltmeter}
For time-series or time-domain measurements (like the measurements of $V_\mathrm{null}$ sampled at a fixed interval $\tau_0$), the magnitude of the Allan deviation (referred to as $\sigma_\mathrm{NVM}$ in the main text), is inversely indicative of the stability of the measurement (i.e. a smaller magnitude indicates greater measurement stability). When plotted on a $\log$-$\log$ scale, the functional dependence of $\sigma_{\mathrm{NVM}}$ on $\tau$ is commonly used to determine the type of noise that dominates a measurement over a given time scale~\cite{allan1987}. For example, the Allan deviation for white noise limited data will scale as $\tau^{-1/2}$ whereas data dominated by $1/f$ noise will have an Allan deviation that appears constant as a function of $\tau$.

NIST special publication 1065~\cite{riley2008} provides extensive details on stability analysis methods for time and frequency domain measurements. The stability and noise analysis protocols used in this work largely followed the guidance of that publication. Specifically, the time series data for $V_\mathrm{null}$ was detrended by subtracting a linear function in time. This removes correlations caused by linear thermal drift~\cite{riley2008}.

 The overlapping Allan variance was calculated using equations (11) from reference~\cite{riley2008}:
\begin{equation}
    \sigma_\mathrm{NVM}^2(\tau) = \frac{1}{2\left(N-2m\right)\tau^2}\sum\limits_{i=1}^{N-2m}\left(\tilde{V}_{\mathrm{null}_{i+2m}}-2\tilde{V}_{\mathrm{null}_{i+m}}+\tilde{V}_{\mathrm{null}_{i}}\right)^2
\end{equation}

Here $\tilde{V}_{\mathrm{null}_{i}}$ is the $i^{\textrm{th}}$ sample of the integrated time series of $V_\mathrm{null}$:

\begin{equation}
    \tilde{V}_{\mathrm{null},i} =\int_0^{i\tau_0}{V_\mathrm{null}dt}
\end{equation}

for $i=1,2,3, \ldots , N$, $m$ is the averaging factor, and $\tau = m\tau_0$ is the general sampling interval. The overlapping Allan variance can increase confidence in stability estimates by utilizing all possible combinations of the data compared to the non-overlapping form~\cite{riley2008}.

\begin{figure}[ht]
   \centering
  \includegraphics[scale=1.2]{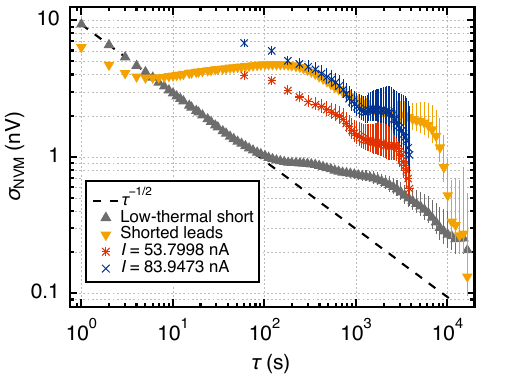}
\caption{Allan deviation comparison of the bare nanovoltmeter and of $V_{\mathrm{null}}$ from direct realizations at $I = 53.8$~nA and $I = 83.9$~nA. Depicted are the overlapping, detrended, Allan deviations calculated for: independent measurements of the nanovoltmeter with a low-thermal short at its input (gray upward-triangles), the same, but with a short at the end of a 1~m long cable (yellow downward-triangles), $V_\mathrm{null}$ measured during direct realization of the ampere  at $I_{GUI} = 53.7889$~nA (red asterisks) and $I_{GUI} = 83.9473$~nA (blue x-symbols). Shown for reference is a $\tau^{-1/2}$ trendline, which is characteristic of noise with a white spectrum. The Allan deviation from the $I_\mathrm{direct}$ data start at $\tau = 60$~s because of the reversal cycles used during ampere realizations. All measurements were performed in temperature- and humidity-controlled laboratories in which temperature fluctuations were less than $\pm$0.5 K. Error bars represent the $k = 1$ Type A uncertainty of the Allan deviation, under the assumption of white noise.} 
   \label{fig.adev}
\end{figure}

To gain a more complete understanding of the factors contributing to our noise floor, we independently measured the Allan deviation $\sigma_\mathrm{NVM}(\tau)$ of the nanovoltmeter itself in a separate laboratory. The second lab had temperature and humidity control similar to that where our unified realization was located. These independent measurements were then compared to Allan deviation data of the null voltage from some of our direct ampere realizations, which used the same nanovoltmeter. The two sets of data are shown as a function of $\tau$ in Fig.~\ref{fig.adev}. For the independent evaluation, we considered two configurations of the nanovoltmeter: (1) one in which the nanovoltmeter terminals were shorted using a low thermal EMF plug that connects directly to the nanovoltmeter's input connector ('Low-thermal short') and (2) one in which we shorted the copper spade terminals together at the end of a 1~m low thermal EMF cable that is furnished with the nanovoltmeter (`Shorted leads'). 

The `Low-thermal short' measurements in Fig.~\ref{fig.adev} show what is likely the minimum achievable noise for the Keithley 2182A nanovoltmeter. For $\tau \lesssim 10^2$~s, this data clearly scale as $\tau^{-1/2}$, indicating a white-noise-dominated spectrum~\cite{allan1987}. For longer times, noise with a different spectra, possibly scaling as $1/f$, can be seen. The `Shorted leads' data is comparable to that of the `Low-thermal short' at short times, exhibiting a white noise spectrum up to $\tau \sim 5$~s. However, this functional behavior quickly transitions to a spectrum more similar to $1/f$ noise where $\sigma(\tau)$ is approximately constant, before eventually ($\tau \sim 2\times10^{2}$~s) transitioning back to a seemingly white spectrum. Thus, just by adding 1~m leads---which were equilibrated for 24 hrs and left undisturbed during measurements---the noise magnitude and spectrum changed substantially. 

Superimposed on the same plot are the nanovoltmeter Allan deviations during measurements of $I_\mathrm{direct}$ at $I = 53.8$~nA and $I= 83.9$~nA. The $I_\mathrm{direct}$ data start at $\tau = 60$~s instead of 1~s because of the reversal cycles used during ampere realizations (i.e., reversals are not needed for the independent nanovoltmeter measurements). We observe that the magnitudes and trends of the $I_\mathrm{direct}$ data are comparable to that of the independent, `shorted leads` arrangement. This suggests that the cabling connecting the nanovoltmeter to devices in the DR may add a sizable amount of noise to our measurements. We should highlight that the variation in magnitude between the $I = 53.8$~nA and $I= 83.9$~nA data is comparable to that between repeated measurements in the independent, `Shorted leads' case. Thus, even though a longer length of cable (internal and external to fridge) was required to obtain $I_\mathrm{direct}$, one should not always expect the Allan deviation to exceed that of the `Shorted leads' case because of the variability between different Allan deviation measurements. The variability is likely due to changes in the lab environment---such as microphonic and/or electromagnetic pickup from adjacent experiments or occupants. We anticipate uncertainty reductions in future campaigns by meticulous design of our experimental wiring.

\providecommand{\noopsort}[1]{}\providecommand{\singleletter}[1]{#1}%

\end{document}